\documentclass[useAMS,usenatbib]{mn2e}
\usepackage[dvips]{graphicx}
\usepackage{longtable}

\begin{document}
 
\title[Origin of lopsidedness in group galaxies] 
{Origin of Disc Lopsidedness in the Eridanus Group of Galaxies}
 
\author[Angiras et al.]{R.A.Angiras$^{1}$\thanks{On leave from St.Joseph's College, Bangalore, India}, C.J.Jog $^2$,A.Omar$^3$,K.S.Dwarakanath$^4$\\
1.School of Pure and Applied Physics,M.G. University, Kottayam, 686560,India \\
2.Department of Physics, Indian Institute of Science, Bangalore, 560012, India \\
3.Aryabhatta Research Institute of Observational Sciences, Manora Peak, Nainital, 263129, India\\
4.Raman Research Institute, Bangalore, 560080, India\\
(E-mails:rangiras{@}rri.res.in,cjjog{@}physics.iisc.ernet.in,aomar{@}aries.ernet.in,dwaraka{@}rri.res.in)}

\date{Accepted.....; Received .....}


\maketitle


\begin{abstract} 
The HI surface density maps for a sample of 18 galaxies in the Eridanus 
group are Fourier analysed. This analysis gives the radial variation of the 
lopsidedness in the HI spatial distribution. The lopsidedness is quantified 
by the Fourier amplitude $A_1$ of the $m=1$ component normalized
to the average value. It is also shown that in the radial region where the stellar 
disc and HI overlap, their $A_1$ coefficients are comparable.
All the galaxies studied show significant lopsidedness in HI. The mean value of $A_1$ in the inner regions of 
the galaxies (1.5 - 2.5 scale lengths) is $\geq 0.2$. 
This value of $A_1$ is twice the average value seen in
the field galaxies. Also, the lopsidedness is found to be smaller for
late-type galaxies, this is opposite to the trend seen in the field
galaxies. These two results indicate a different physical origin for 
disc lopsidedness in galaxies in a group environment compared to
the field galaxies. Further, a large fraction ($\sim$ 30\%) shows a higher 
degree of lopsidedness ($A_1 \geq 0.3$). It is also seen that
the disk lopsidedness increases with the radius as demonstrated in earlier
studies, but over a radial range that is two times
larger than done in the previous studies. 
The average lopsidedness of the halo potential is 
estimated to be $\sim 10$\%, assuming that the lopsidedness in HI disc 
is due to its response to the halo asymmetry.
\end{abstract}
 
 
\section{INTRODUCTION} 

The presence of non-axisymmetric distribution of atomic hydrogen gas (HI) in spiral galaxies has been known for many
years. In a pioneering study, Baldwin et al. (1980)  pointed out large scale spatial asymmetry in the HI gas
distribution of four nearby spiral galaxies M 101, NGC 891, NGC 2841 and IC 342. All these galaxies show HI gas
extending out to larger radii on one side than on the other.

Since then the non-axisymmetric distribution of HI gas has been deduced for much larger samples by studying the
asymmetry in the global HI profiles (Richter \& Sancisi 1994, Matthews et al. 1998, Haynes et al. 1998). These studies
revealed that the HI profiles on the receding and approaching sides are asymmetric. It was inferred from the sample
that at least 50\% of galaxies studied show HI lopsidedness. However, without the analysis of 2-D maps of HI discs, the above studies
can only indicate the result of lopsidedness caused jointly by the spatial and velocity distribution. Such a quantitative measurement of HI spatial asymmetry from 2-D maps is still to be carried out. 

A large fraction of galaxies show asymmetry. This indicates that the lopsidedness is  sustainable over a long period of time.
Yet it's physical origin is not clearly understood. The cause of disc lopsidedness has been attributed to a variety of
physical processes, such as the disc response to halo lopsidedness which could arise due to tidal interactions (Jog,
1997), or due to mergers with satellites (Zaritsky \& Rix 1997), or asymmetric gas accretion (Bournaud et al. 2005). The asymmetry can also be generated 
due to a disc offset in a spherical halo \citep{Noord01}.Thus a study of HI asymmetry in the outer parts as done in this paper can
 give a direct handle on the halo asymmetry if the disc lopsidedness
 arises due to halo asymmetry.

The existence of asymmetry in the velocity domain, i.e., kinematical lopsidedness, has also been detected in spiral
galaxies.These studies are based on the analysis of asymmetry of the rotation curves on the approaching and receding
sides of a galaxy \citep{Swaters99} and also by analysing the HI velocity field of a spiral galaxy directly
\citep{Schoenmakers97}. It is postulated that the same perturbation potential that gives rise to spatial lopsidedness
will also unavoidably give rise to kinematical lopsidedness (Jog 1997, Jog 2002).

With the advent of near-IR observations in recent years, spatial lopsidedness has also been detected in the
distribution of old stellar population that make up the main mass component of galactic discs (Block et al. 1994, Rix
\& Zaritsky 1995). The harmonic analysis is used to Fourier analyse the photometric data on the galaxy images, which
gives a quantitative measurement of spatial lopsidedness as a function of radius. It is found that more than 30\% of
galaxies show strong spatial lopsidedness in near-IR(Rix \& Zaritsky 1995, Zaritsky \& Rix 1997, Bournaud et al. 2005);
but the increased sky brightness in the near-IR bands limits the measurement of Fourier components of stellar asymmetry
to radii less than 2.5 exponential disc scale lengths. It is also not known whether the lopsidedness in HI and near-IR bands are correlated.

In the present work, we Fourier-analyse the HI surface density distribution for a sample of eighteen galaxies in the
Eridanus group of galaxies (Omar \& Dwarakanath 2005a). To our knowledge, this is the first time that an analysis
of this kind has been applied for quantitative measurement of HI spatial asymmetry. The availability of interferometric
2-D maps of galaxies has made this study possible. Since the HI gas usually extends farther out than the stars, the
disc lopsidedness can be measured up to twice or more the radial distance than it was possible using stellar light.
Since the lopsidedness is observed to increase with radius (Rix \& Zaritsky 1995, Bournaud et al. 2005), the Fourier
amplitude measured at these distances are expected to provide better constraints on the generating mechanisms for disc
lopsidedness. In addition to this we have also compared the lopsidedness in HI with that in the near-IR 
band, and show these to be comparable. We show that the main physical
mechanism for the origin of disc lopsidedness for the group galaxies is
different than for the field galaxies.

This paper is organized as follows. In Section 2, the HI data used for this analysis is described. The harmonic
analysis method and the results derived from HI maps and R-band images are presented in Section 3. A few general points
and the origin of disc lopsidedness are discussed in Section 4. Section 5 contains a brief summary of the conclusions from the paper.

\section{Data: The Eridanus Group of Galaxies} 

The Eridanus group is a loose group of galaxies at a mean distance of $\sim23\pm 2$ Mpc in the southern hemisphere
(\hbox{$\sim 3^{h}\leq\alpha\leq 4^{h}30^{\prime}$}, $\sim -10^{\circ}\geq\delta\geq-30^{\circ}$).  From the redshift
values, $\sim 200$ galaxies are associated with this group with heliocentric velocities in the range of $\sim
1000-2200$ km s$^{-1}$. The observed velocity dispersion is $\sim 240$ km s$^{-1}$.

Even though there are sub-groups within the system, the overall population mix of the galaxies in the Eridanus group
was found to be $\sim30\%$ ellipticals and lenticulars and $\sim70\%$ spirals and irregulars (Omar \& Dwarakanath
2005a).Though HI was detected in 31 galaxies out of the 57 selected for observation by Omar and Dwarakanath using the
Giant Meter-wave Radio Telescope (GMRT), the spiral galaxies under consideration here form a subset of these. These spiral galaxies were chosen on the basis of their inclination, with inclinations in the range of $20^{\circ}$ and $80^\circ$. This
criterion was adopted so as to get  good velocity and surface density maps.

If a galaxy is almost face on, the circular velocity information derived from the velocity map of the galaxy will be
reduced to a great extent resulting in greater uncertainty in the inclination. Similarly, if a galaxy is viewed edge-on, the
line of sight velocity information will be of good quality, but the surface density map will not be suitable for the spatial
lopsidedness analysis. With these criteria, an inclination range of $20^\circ$-$80^\circ$ was found suitable.

In addition to this, we eliminated those galaxies where the detection in HI was patchy such as NGC 1415.
The positions and Heliocentric velocities of selected galaxies are given in Table 1.

\begin{table*}
\caption{Type, position, inclination,position angle and scale lengths of observed galaxies}
\noindent
\begin{tabular}{@{}llrrrcccccc@{}}
\hline
\hline
Galaxy        &	  Type&\bf{$\alpha$}\small(J2000) &\bf{$\delta$}\small(J2000)&\bf{$cz$}&Inclin(i)&    PA& $D_{H}$&$D_{R25}$&$R_{J}$&$R_{K}$\\
              &       &~h~~m~~s~ & ~~\hbox{$^\circ$}~~$'$~~$''$~& (km/s)&	(Deg)  & (Deg)&   (kpc)&     (kpc)&    (kpc)&   (kpc)\\
\hline
              &	      &          &	   &	&	   &      &         &	       &	 &	 \\
NGC 1309      &  SAbc &03~22~06.5&-15~24~00&2135&        20&   210&     24.1&      13.4&    1.42 &    1.32\\
UGCA 068      & SABcdm&03~23~47.2&-19~45~15&1838&        34&    35&     16.9&      10.3&    1.29 &    1.08\\
NGC 1325      &  SAbc &03~24~25.4&-21~32~36&1589&        71&   232&       44&      38.5&    4.6  &    4.61\\
NGC 1345      &  SBc  &03~29~31.7&-17~46~40&1529&        34&    88&     23.9&      8.51&    0.93 &    1.15\\
NGC 1347      &  SBcd &03~29~41.8&-22~16~45&1759&        26&   328&     12.9&      8.51&    1.46 &    1.44\\
UGCA 077      &  SBdm &03~32~19.2&-17~43~05&1961&        66&   149&     22.1&      12.1&    *    &       *\\
IC 1953       &   SBd &03~33~41.9&-21~28~43&1867&        37&   129&     21.3&      24.2&     3.78&    4.19\\
NGC 1359      &  SBcm &03~33~47.7&-19~29~31&1966&        53&   325&        *&      17.9&    2.67 &       *\\
NGC 1371      &  SABa &03~35~02.0&-24~55~59&1471&        49&   136&       61&      34.5&    3.34 &    3.37\\
ESO 548- G 049&  S?   &03~35~28.1&-21~13~01&1510&        71&   128&     14.9&      6.27&     1.18&       *\\
ESO 482- G 013&	Sb    &03~36~53.9&-24~54~46&1835&        63&    65&     12.9&	   7.17&     0.65&   0.74 \\
NGC 1385      &  SBcd &03~37~28.3&-24~30~05&1493&        40&   181&     19.8&      30.5&    3.21 &    2.98\\
NGC 1390      &  SB0/a&03~37~52.2&-19~00~30&1207&        60&    24&     15.2&      8.96&    0.74 &    0.87\\
NGC 1414      &  SBbc &03~40~57.0&-21~42~47&1681&        80&   357&     19.9&      11.2&    1.96 &    1.81\\
ESO 482- G 035&  SBab &03~41~15.0&-23~50~10&1890&        49&   185&     16.4&      14.8&    1.95 &   1.96\\
NGC 1422      &  SBab &03~41~31.1&-21~40~54&1637&        80&    65&     15.4&      18.4&    2.44 &    1.84\\
MCG -03-10-041&  SBdm &03~43~35.5&-16~00~52&1215&        57&   343&     19.2&     13.9 &   3.13  &    2.36\\
ESO 549- G 035&  Sc   &03~55~04.0&-20~23~01&1778&        56&    30&     14.7&       *  &     *   &    	*\\
\hline
\hline
\end{tabular}
\label{Table 1}
\end{table*}
The HI surface density and velocity maps of the selected galaxies used in this analysis were derived out of image cubes
which were convolved to a common resolution of $20^{\prime\prime}\times 20^{\prime\prime}$. A $3\sigma$ column density
sensitivity of $10^{20}$~cm$^{-2}$ was obtained for $20^{\prime\prime}$ resolution surface density images. The velocity
resolution was $\sim 13.4$~km s$^{-1}$. A typical HI surface density contour map and velocity contour map superposed on
the DSS image are shown in Fig.1.

\begin{figure*}
\includegraphics[width=100mm,height=60mm]{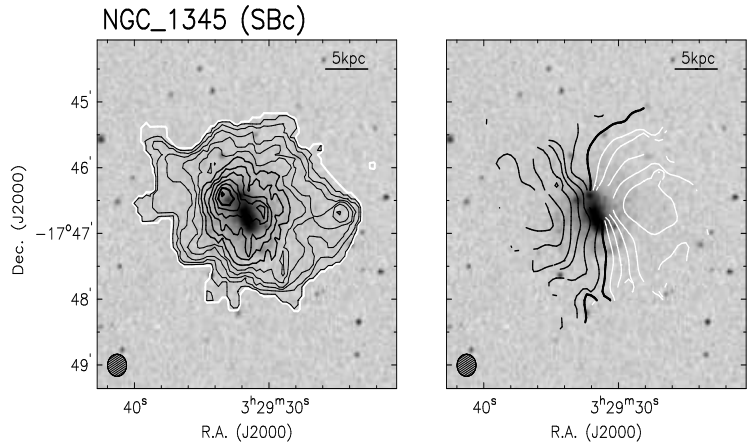}

\caption{Typical HI and velocity map of a galaxy in the Eridanus group. The contours are superposed on DSS gray scale
image. The beam is shown at the bottom left hand corner.The surface density map contour levels are separated by
$2\times10^{20}$~cm$^{-2}$. The first contour, shown in white, is at a column density of $1\times10^{20}$~cm$^{-2}$.The
first thick contour begins at $1.2\times10^{21}$~cm$^{-2}$ and the second one is at $2.2\times10^{21}$~cm$^{-2}$.  The
velocity contours for the approaching side are shown in white while that for the receding side are shown in black. The
velocity contours differ from each other by $10$~km s$^{-1}$. The thick dark line near the centre of the galaxy denotes
the systemic velocity (Omar \& Dwarakanath 2005a).}

\end{figure*}

With a view to studying the spatial lopsidedness of the stellar component, the corresponding R-band images of these galaxies were analysed. The R-band images were obtained from Aryabhatta Research Institute of observational-sciencES (ARIES)using the 104 cm Sampurnanand telescope. The details of the observations and data reduction methods are given elsewhere(Omar \& Dwarakanath,2006).The final re-gridded images had a typical resolution of $1^{\prime\prime}.0\times 1^{\prime\prime}.0$ and a limiting surface brightness of $\sim 26.0$ mag arcsec$^{-2}$.

\section{ Harmonic Analysis}
\subsection{Spatial lopsidedness of HI images}

The HI maps are Fourier analysed to study the spatial asymmetry of the HI surface density distribution with the
help of velocity maps. The HI surface density is extracted along the concentric annuli each of width $20^{\prime\prime}$
after correcting for the projection effects. Such an analysis is prompted by the assumption that, an ideal galaxy can
be assumed to be made up of a set of concentric rings along which the gas is rotating about the galactic centre
(Begeman,1989). Though in general these rings are elliptical, the departure from circular symmetry is very small and
hence they can be assumed to be circular rings.
 
From the geometry of the problem, at each radius($R$) in the velocity map,

\begin{eqnarray}
V(x,y)&=&V_{0}+V_{c}\cos(\phi)\sin(i) \nonumber\\
      & & +V_{exp}\sin(\phi)\sin(i) 
\end{eqnarray}

where
\begin{eqnarray}
\cos(\phi)=\frac{-(x-x_{0})\sin(PA)+(y-y_{0})\cos(PA)}{R}\nonumber
\end{eqnarray}
\begin{eqnarray}
\sin(\phi)=\frac{-(x-x_{0})\cos(PA)+(y-y_{0})\sin(PA)}{R\cos(i)}\nonumber
\end{eqnarray}
\begin{eqnarray}
R=\sqrt{(x-x_{0})^{2}+\frac{(y-y_{0})^2}{\cos(i)^2}}
\end{eqnarray}

Here, $V(x,y)$ is the observed radial velocity at the rectangular sky coordinate $(x,y)$, $V_{0}$, heliocentric
recession velocity and $V_{c}$, the circular velocity.$V_{exp}$ is the expansion velocity which in this case was taken
to be zero. $\phi$ is the azimuthal angle measured in the anti-clockwise direction in the plane of the galaxy and
$(x_{0},y_{0})$ is the dynamical centre of the galaxy. In all the calculations the position angle ($PA$), was measured
in the anti-clockwise direction from the north to the receding half of the galaxy. Using these equations and the
velocity maps, the Groningen Image Processing System (GIPSY) routine ROTCUR was used by Omar \& Dwarakanath (2004a) to
determine the five unknown quantities,$(x_{0},y_{0})$, $V_{0}$,$V_{c}$,$PA$ and the inclination ($i$) in an iterative
manner described by \citep{Begeman89}. These values are used here to derive the geometrical parameters used in HI-map
harmonic analysis.

The important step in this procedure was determining the dynamical centre of the galaxy about which the gas in each of
the rings is assumed to be rotating and holding it fixed. If the centre and the systemic velocity were not held fixed
for the outer rings,i.e., if they were allowed to wander about, the resulting $2^{nd}$ harmonic coefficients from the
velocity map analysis tended to rearrange themselves so as to minimize the effects of
lopsidedness\citep{Schoenmakers97}. The centre fixing and calculation of systemic velocity was carried out as per the
technique prescribed by Begeman (1989). The optical centre and optical velocity were given as the initial guess. PA and
inclination derived from an elliptical fit to the optical isophotes were held fixed. This fixed the centre. In all the
galaxies which were included in this study, the dynamic centre was very close to the optical centre. The systemic
velocity was taken to be the mean value for each of the rings. The PA and inclination were determined as per the
description given in Omar \& Dwarakanath (2005a).

To derive the surface density harmonic coefficients as well as the velocity harmonic coefficients, these values as well
as the surface density and velocity maps were given to the GIPSY task RESWRI, a task which is an offshoot of ROTCUR
based on the harmonic analysis idea developed by Schoenmakers et al.(1997). The radii of each of the rings were
separated by $10^{\prime\prime}$ and the width of each ring was $20^{\prime\prime}$. To avoid beam smearing along the
minor axis, a cone of $10^{\circ}$ about the minor axis of the galaxy was not included in the analysis. A uniform
weightage was given for each of the points within one annulus.At each of these rings, with the same values of PA and
inclination derived from the velocity map analysis, the surface density values were extracted from the surface density
map, which were Fourier expanded. The programme parameters were set so as to return ten Fourier harmonic coefficients
in the velocity and spatial domain. The resulting harmonic coefficients were recast so that the surface density could
be modelled as

\begin{equation}
\sigma(R,\phi)=\sigma_{0}(R)+\sum_{m}a_{m}(R)\cos (m\phi-\phi_{m}(R))  
\end{equation}

Here $\sigma_{0}(R)$ is the mean surface density at a given radius $R$. $\phi$ is the azimuthal angle in the plane of
the galaxy and $\phi_{m}$ is the phase of the $m^{th}$ Fourier coefficient.  From this analysis, the harmonic
coefficients $a_{1}(R)$ were extracted out and the normalized harmonic coefficients $A_{1}(R)=a_{1}(R)/\sigma_{0}(R) $
was calculated for various rings. $\phi_{1}(R)$ for various rings were also determined. The errors in each of these
coefficients were determined assuming that the coefficients $a_{1}(R),\sigma_{0}(R),\phi_{1}(R)$ to be independent of
each other. This procedure was adopted so as to easily compare the $A_{1},\phi_{1}$ values with the values derived from
a similar analysis of the near-IR data for stars (Rix \& Zaritsky 1995, Bournaud et al. 2005).

The type of the galaxy, its mean inclination, mean position angle, HI diameter, optical diameter and the scale lengths
used (K and J bands) are taken from Omar \& Dwarakanath (2006) and are tabulated (Table 1). The resulting
$A_1$ and $\phi_1$ vs. $R$ are plotted with the radius scaled in terms of the near-IR scale lengths $R_K$ or $R_J$ as
shown in Figures 2 and 3 respectively. This scaling was done in order to facilitate a comparison with the near-IR
values of stellar asymmetry over the same radial range as given in the literature. In the case of some of the Eridanus
group of galaxies, where no scale lengths were available (denoted by $*$ in the last two columns in Table 1), a mean
scale length of $\sim 2$kpc was taken. The mean values for $A_{1}$ obtained over the range of 1.5 to 2.5 stellar
exponential disc scale lengths ( $R_K$ values are preferred) are given in column 6 in Table 2.
The measured values of lopsidedness are much higher than that in the field galaxies. This will be discussed in detail in Section 4.

\begin{figure*}
\includegraphics[width=84mm,height=25mm]{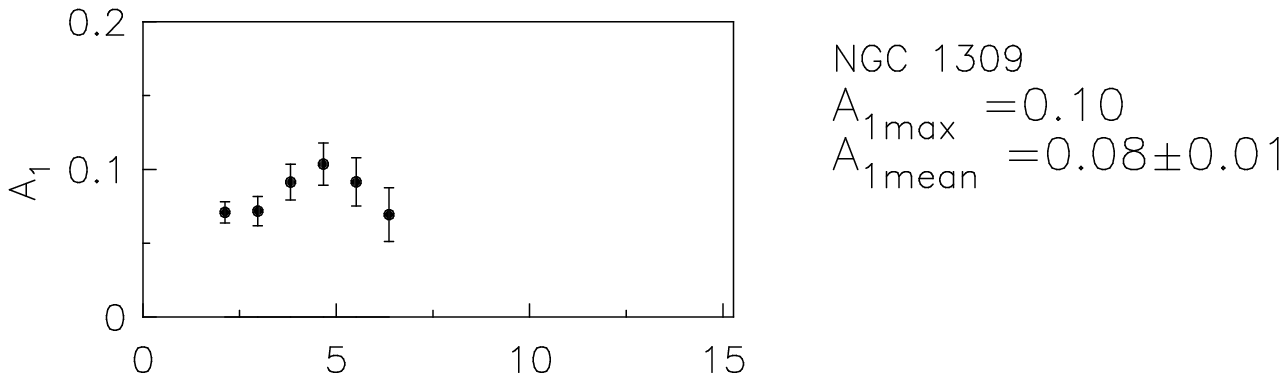}
\includegraphics[width=84mm,height=25mm]{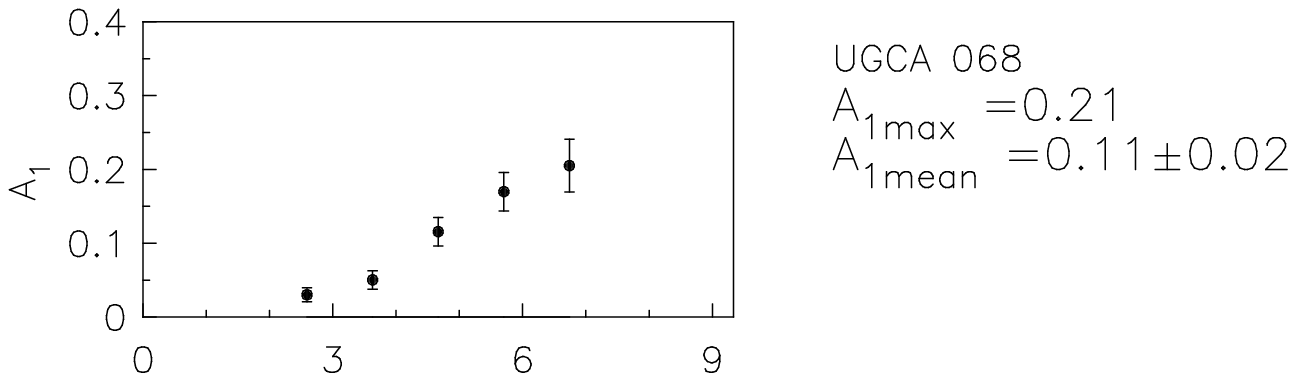}
\includegraphics[width=84mm,height=25mm]{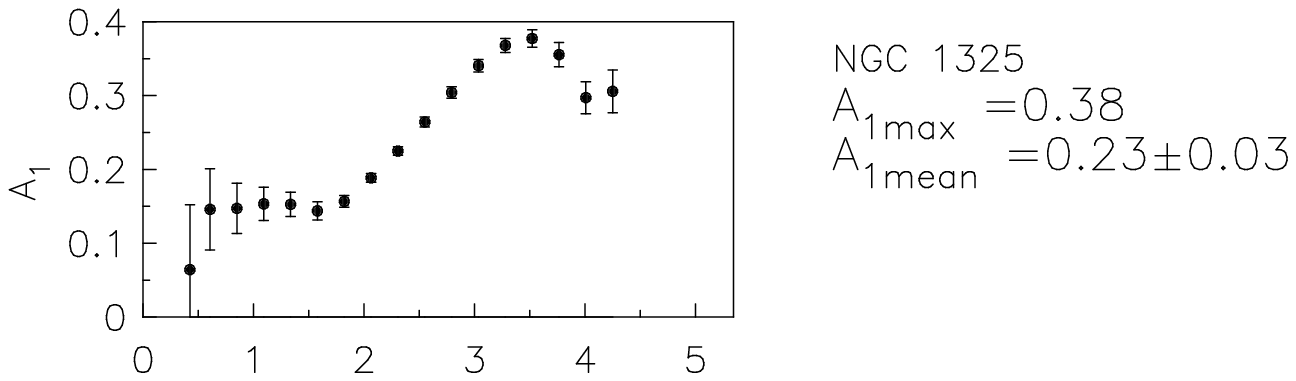}
\includegraphics[width=84mm,height=25mm]{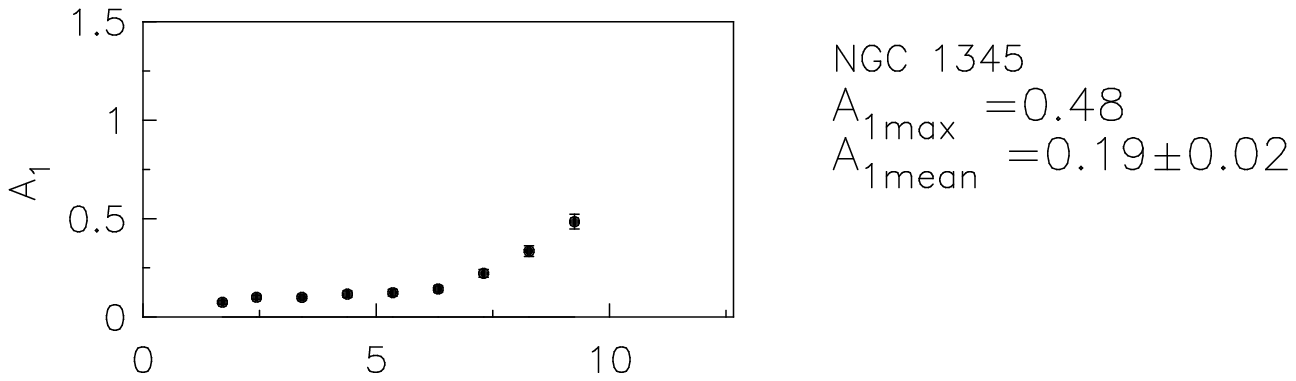}
\includegraphics[width=84mm,height=25mm]{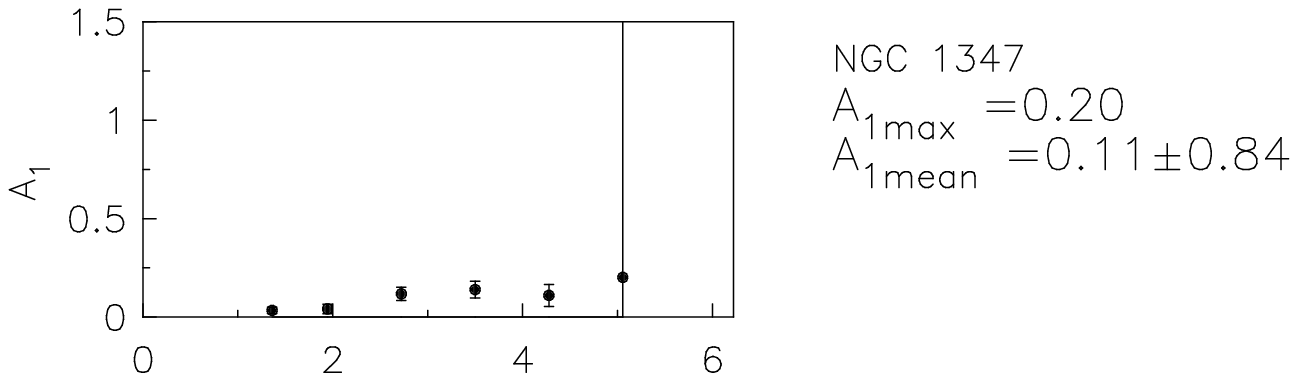}
\includegraphics[width=84mm,height=25mm]{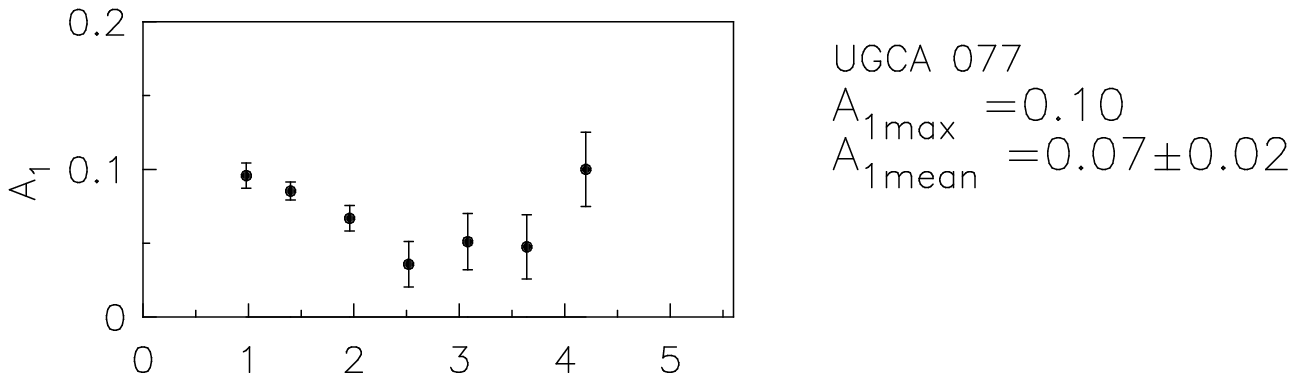}
\includegraphics[width=84mm,height=25mm]{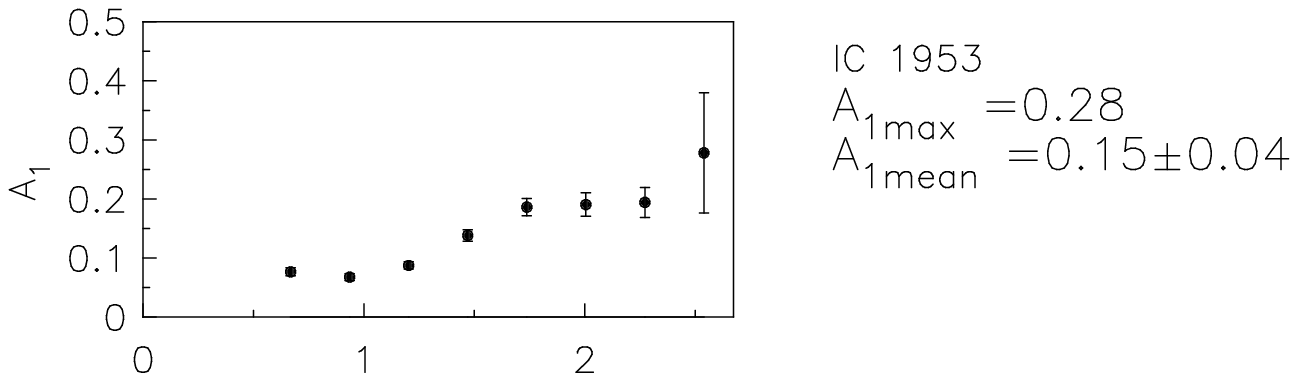}
\includegraphics[width=84mm,height=25mm]{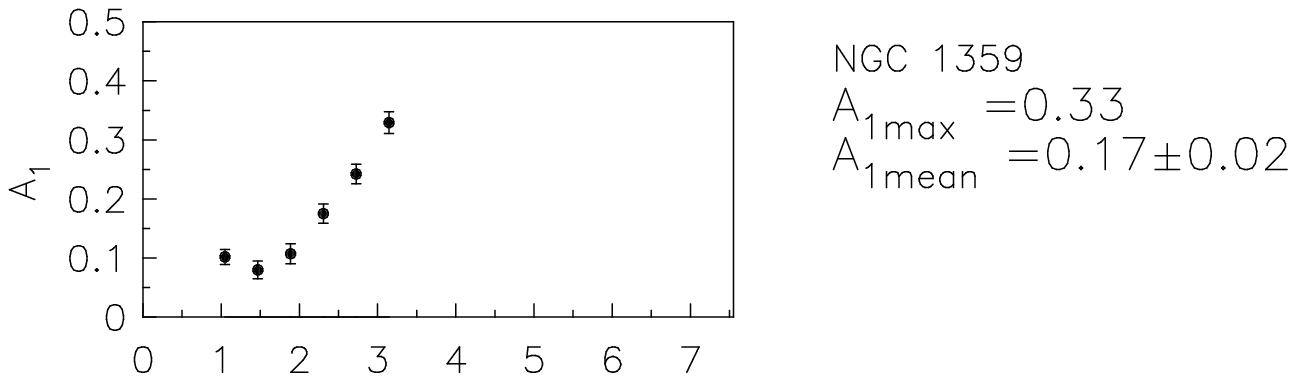}
\includegraphics[width=84mm,height=25mm]{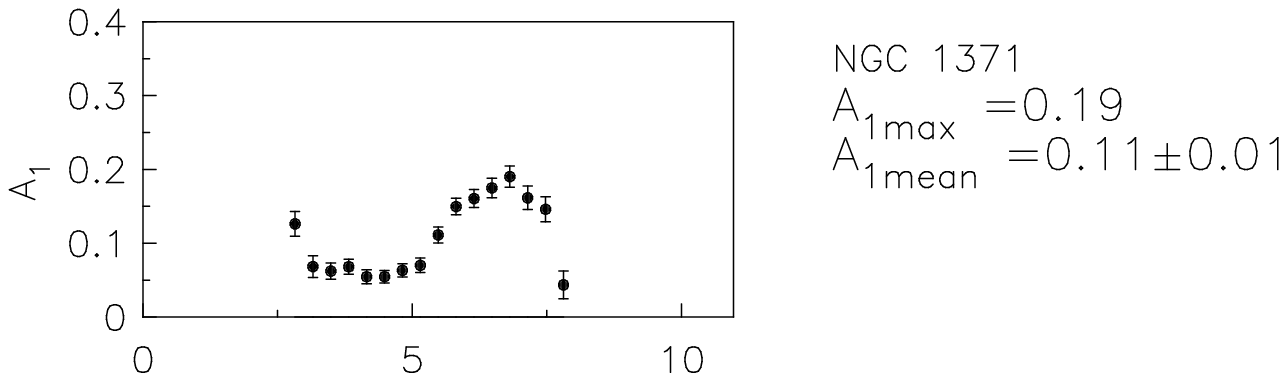}
\includegraphics[width=84mm,height=25mm]{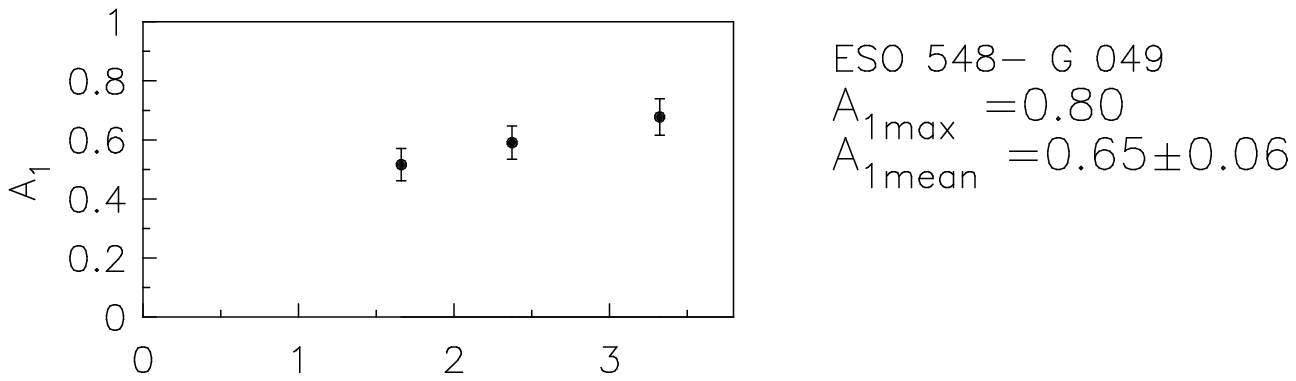}
\includegraphics[width=84mm,height=25mm]{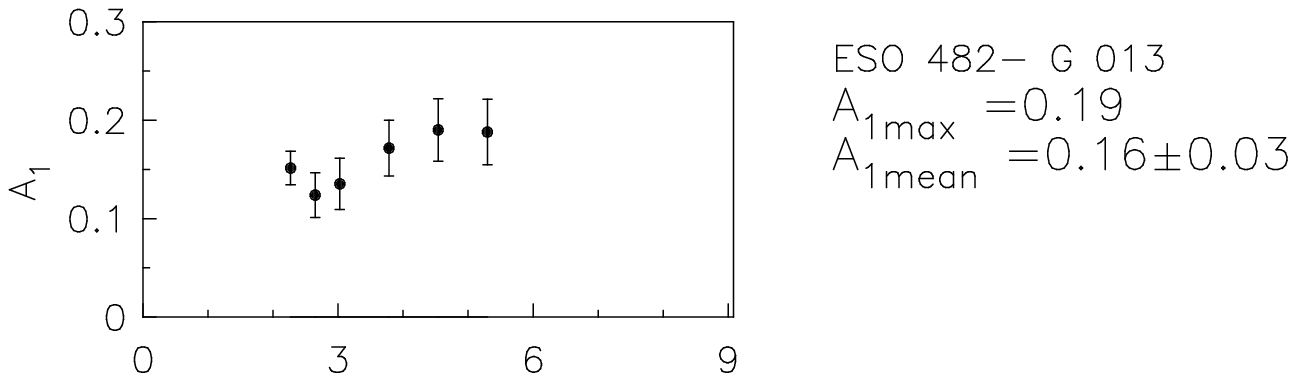}
\includegraphics[width=84mm,height=25mm]{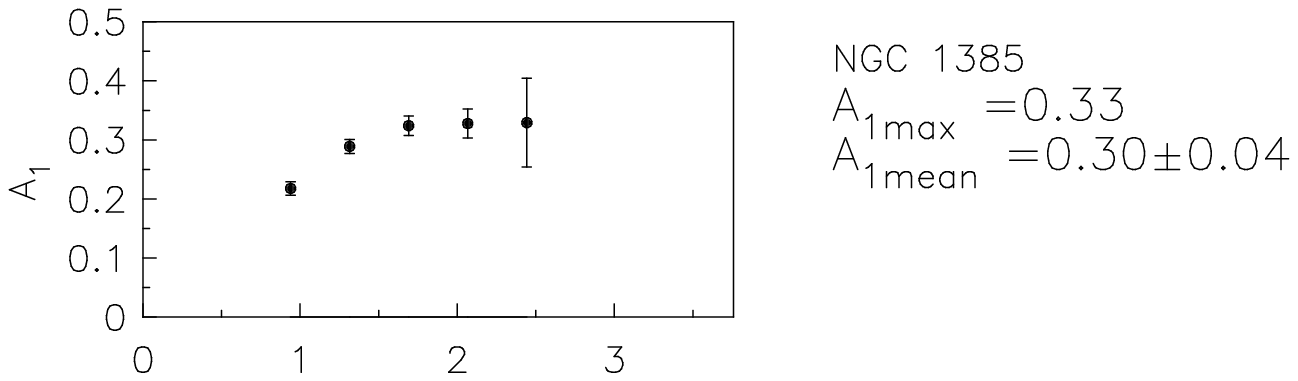}
\includegraphics[width=84mm,height=25mm]{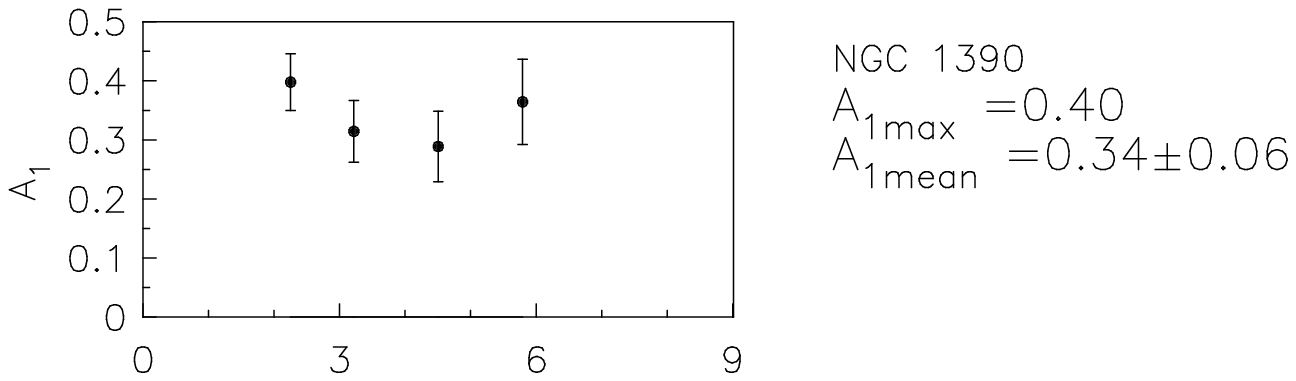}
\includegraphics[width=84mm,height=25mm]{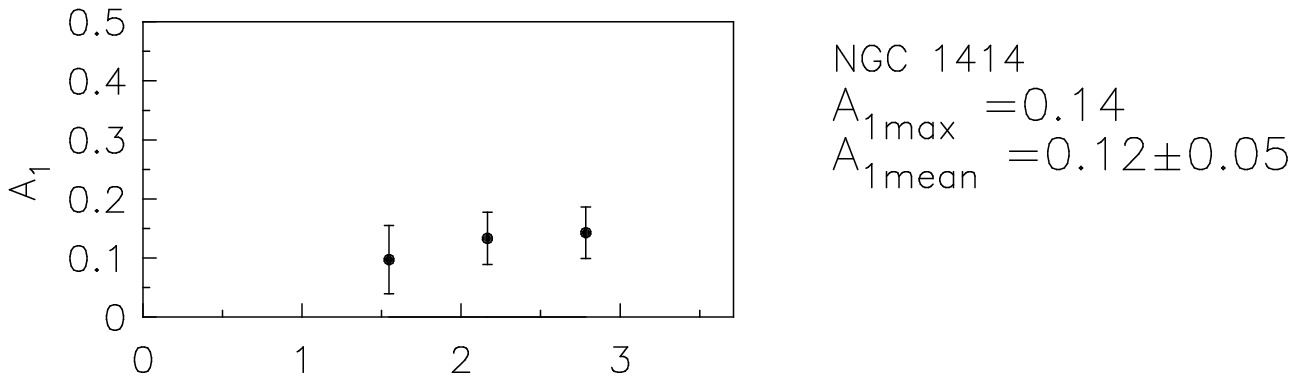}
\includegraphics[width=84mm,height=25mm]{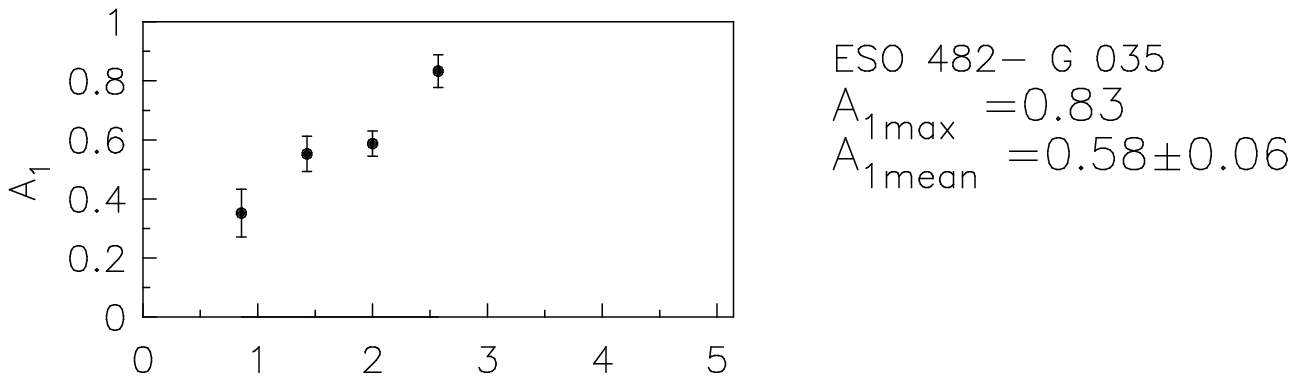}
\includegraphics[width=84mm,height=25mm]{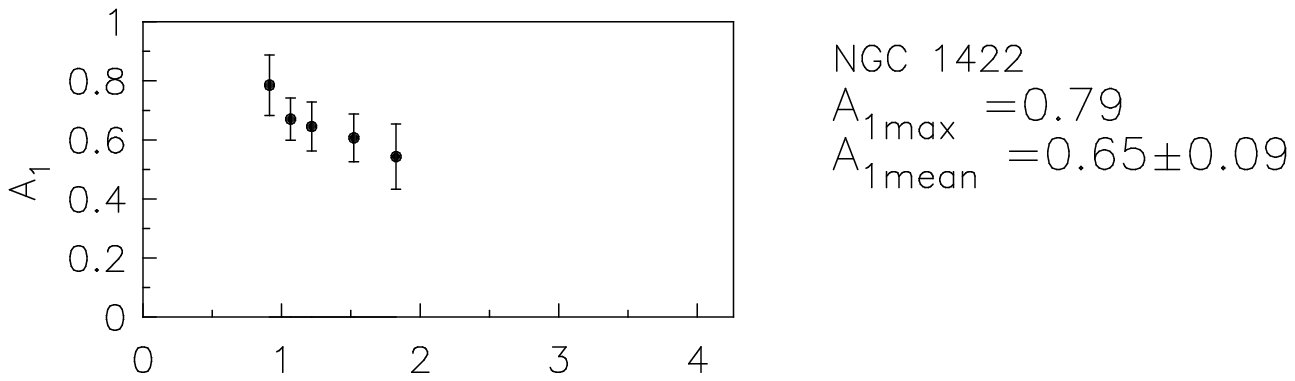}
\includegraphics[width=84mm,height=25mm]{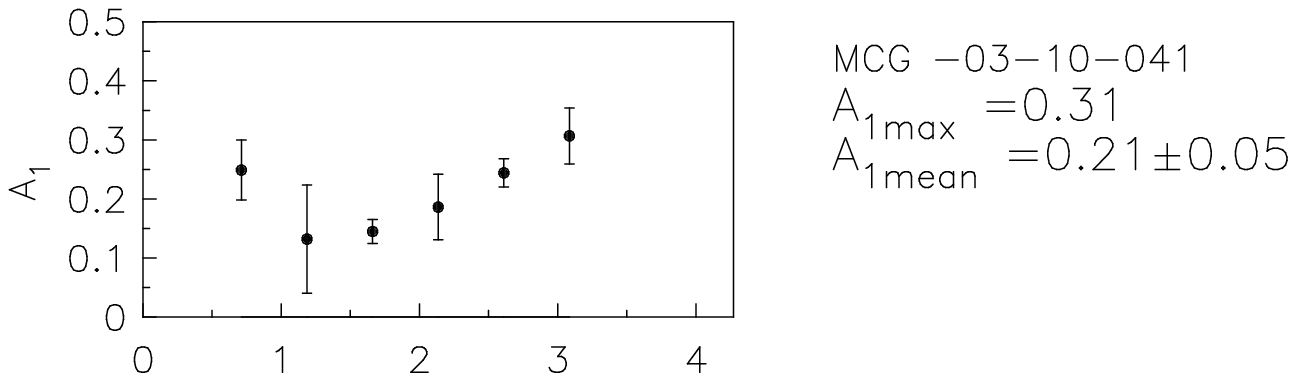}
\includegraphics[width=84mm,height=25mm]{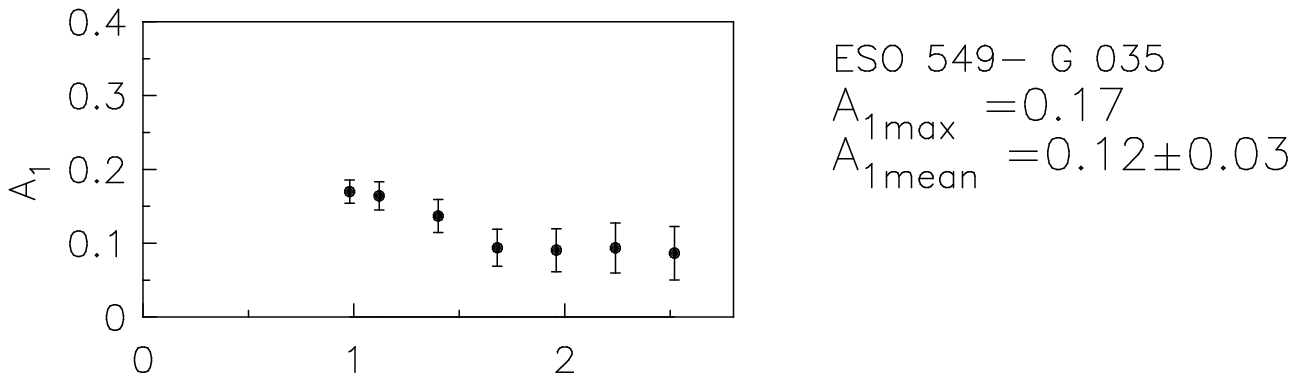}

\caption{$A_{1}$ coefficients of galaxies in the Eridanus group as a function of $R/R_{K}$.For NGC 1359 and ESO 548 -G
049 the J band scale length is used. For UGCA 077 \& ESO 549 -G 035, a scale length of 2kpc is used.Here $A_{1mean}$ is
the mean value of $A_{1}$ over the whole HI disc}

\end{figure*}

\begin{figure*}
\includegraphics[width=84mm,height=25mm]{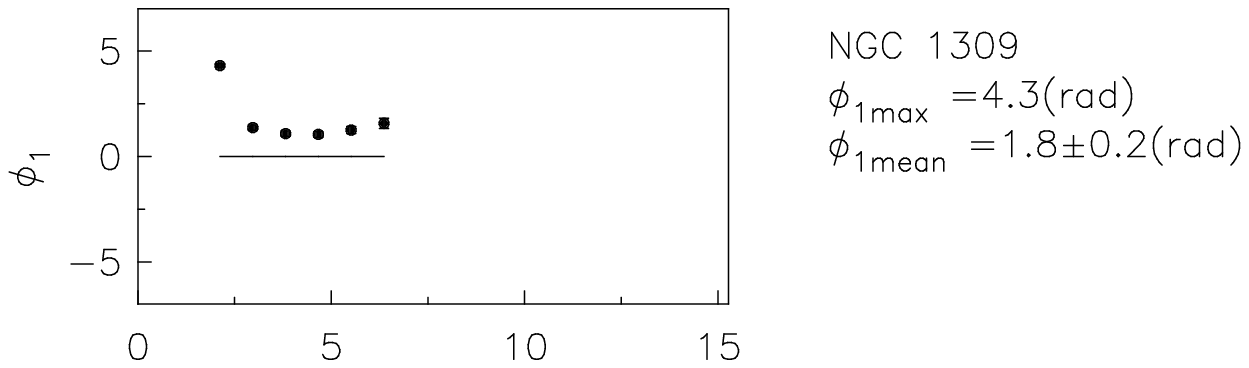}
\includegraphics[width=84mm,height=25mm]{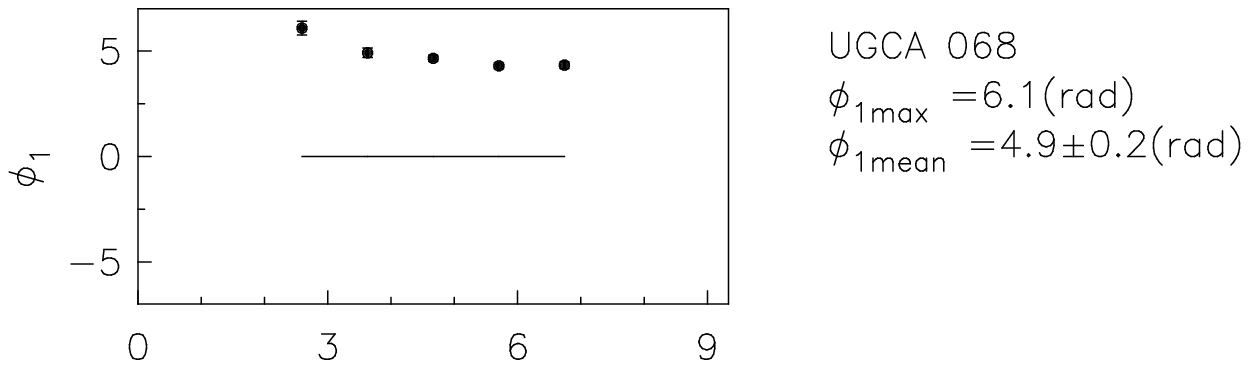}
\includegraphics[width=84mm,height=25mm]{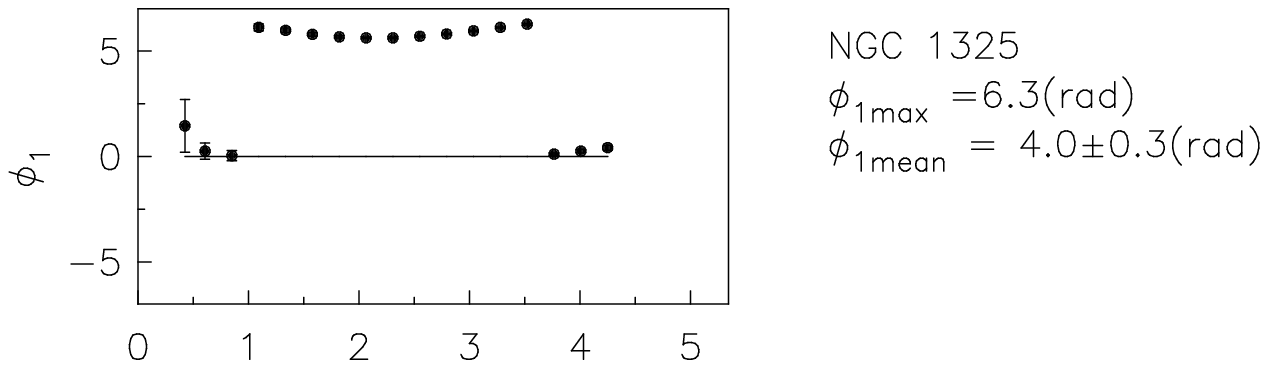}
\includegraphics[width=84mm,height=25mm]{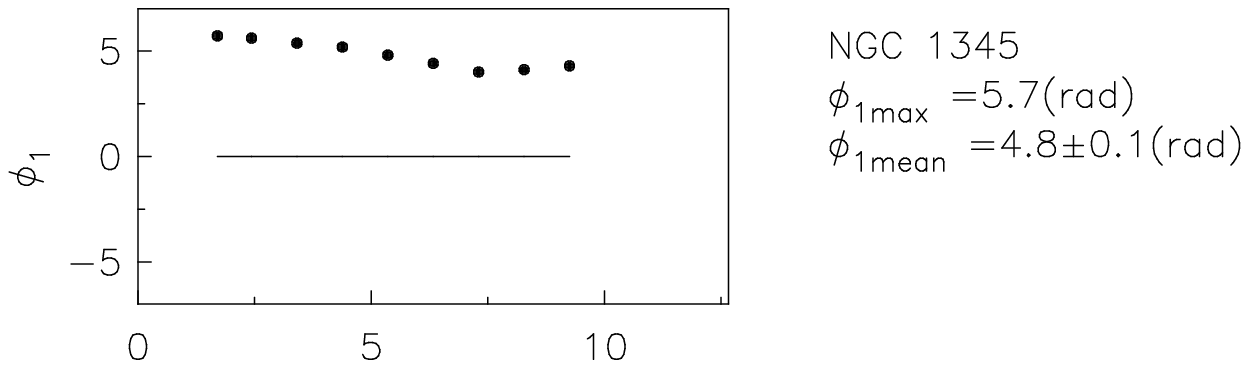}
\includegraphics[width=84mm,height=25mm]{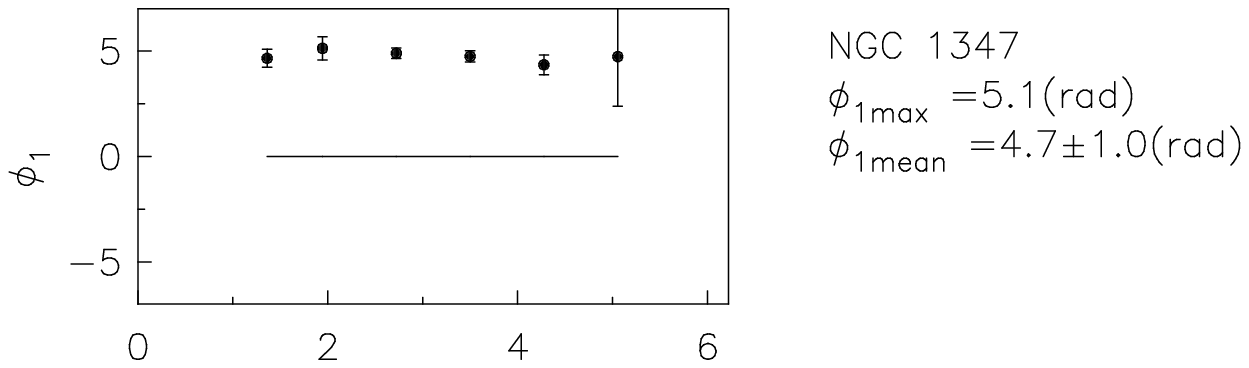}
\includegraphics[width=84mm,height=25mm]{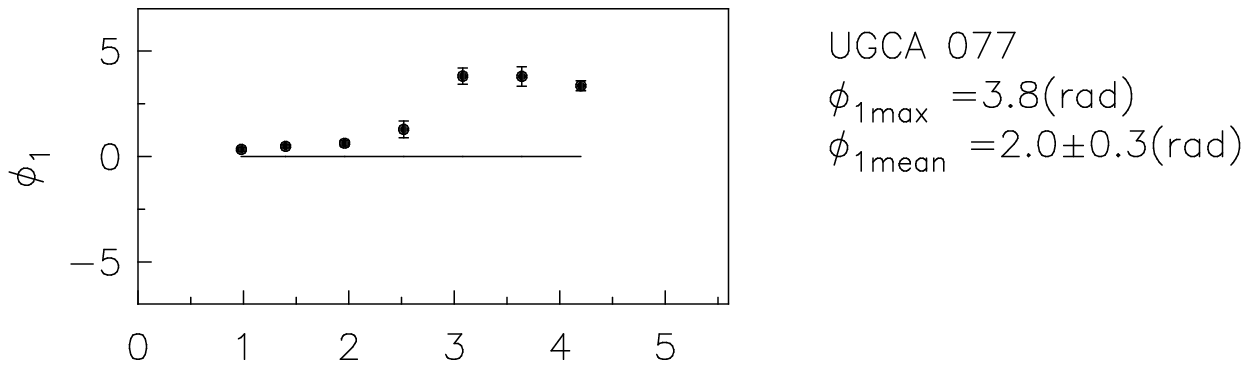}
\includegraphics[width=84mm,height=25mm]{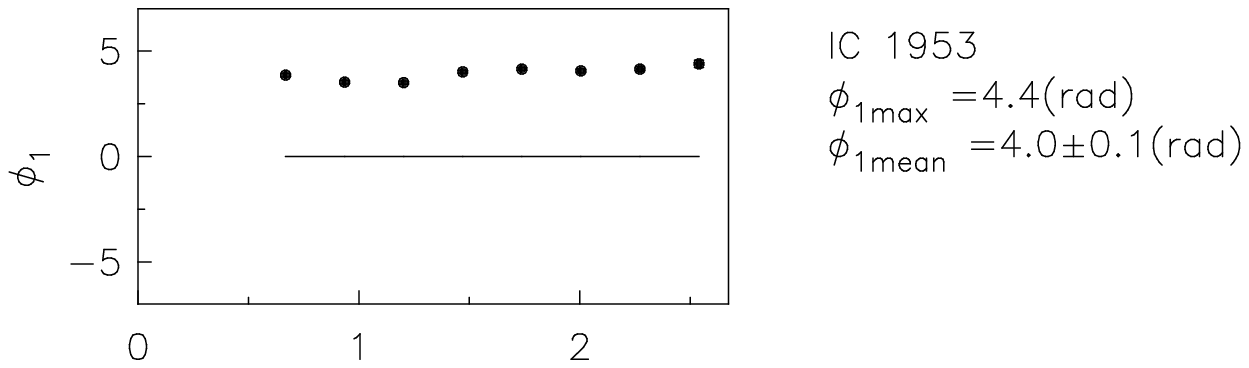}
\includegraphics[width=84mm,height=25mm]{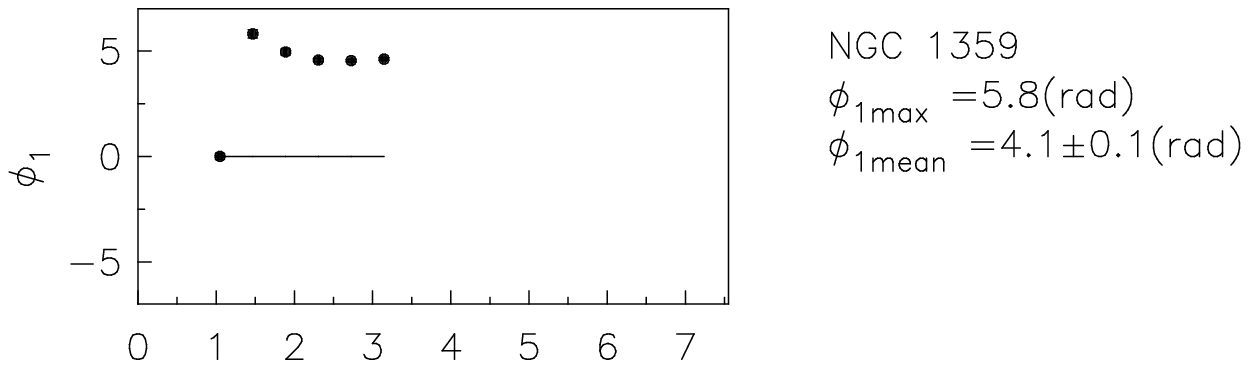}
\includegraphics[width=84mm,height=25mm]{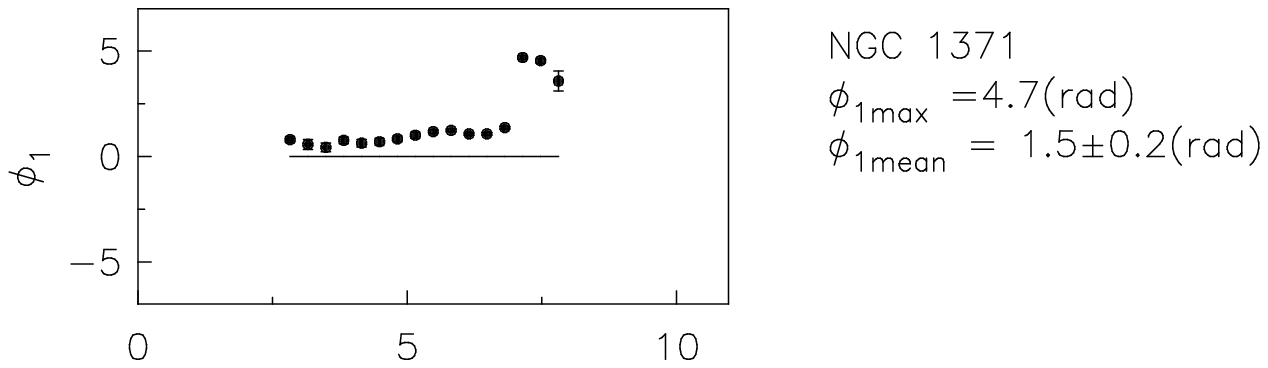}
\includegraphics[width=84mm,height=25mm]{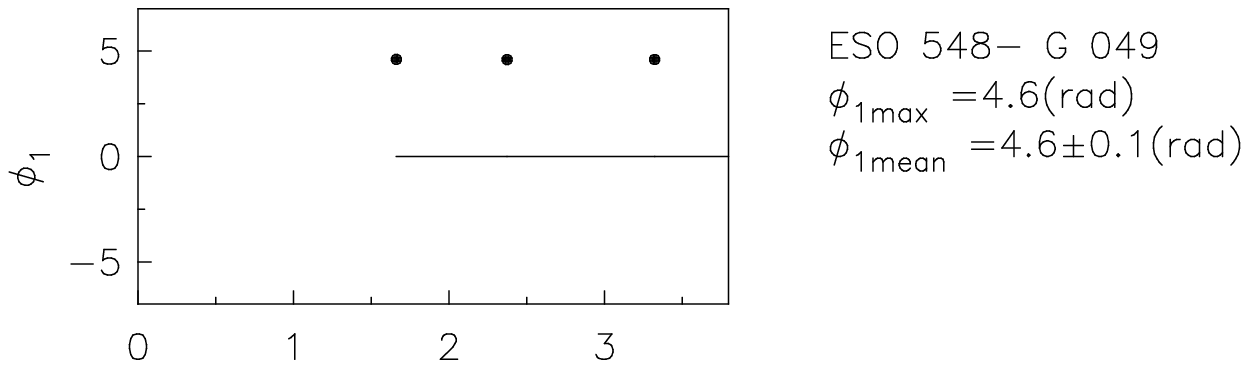}
\includegraphics[width=84mm,height=25mm]{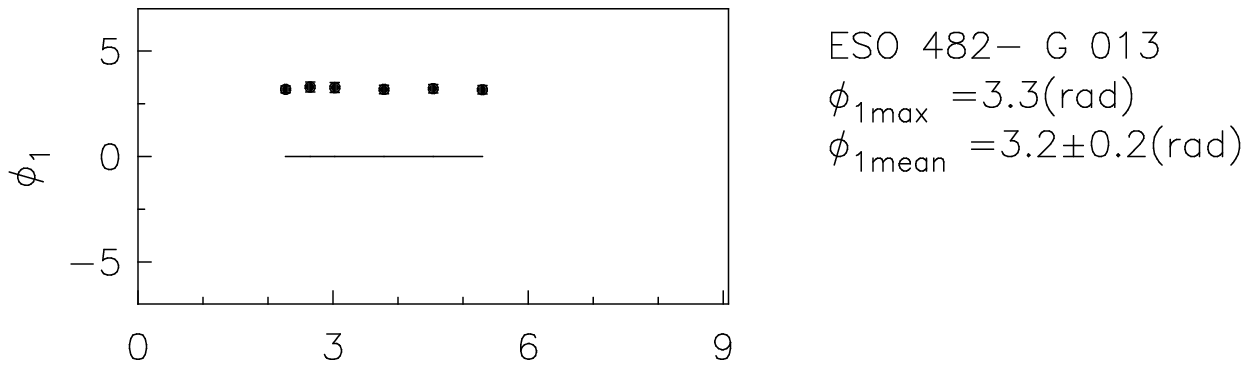}
\includegraphics[width=84mm,height=25mm]{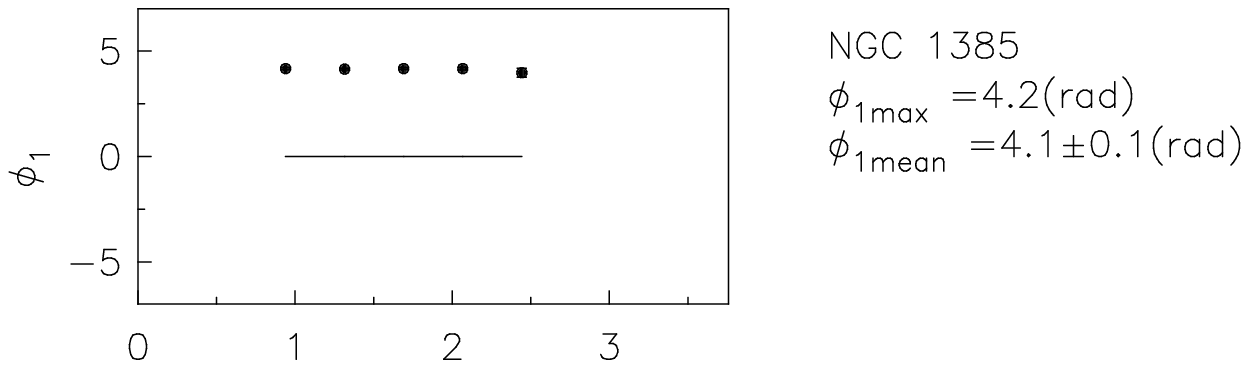}
\includegraphics[width=84mm,height=25mm]{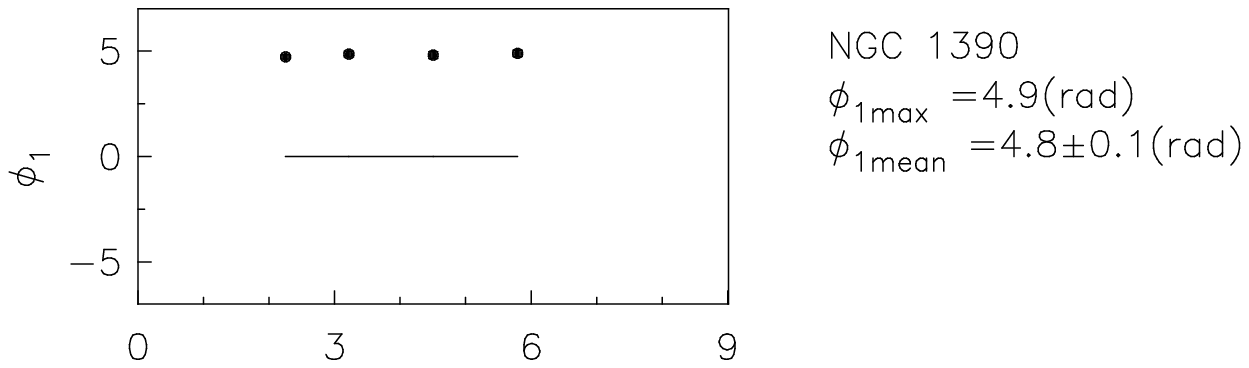}
\includegraphics[width=84mm,height=25mm]{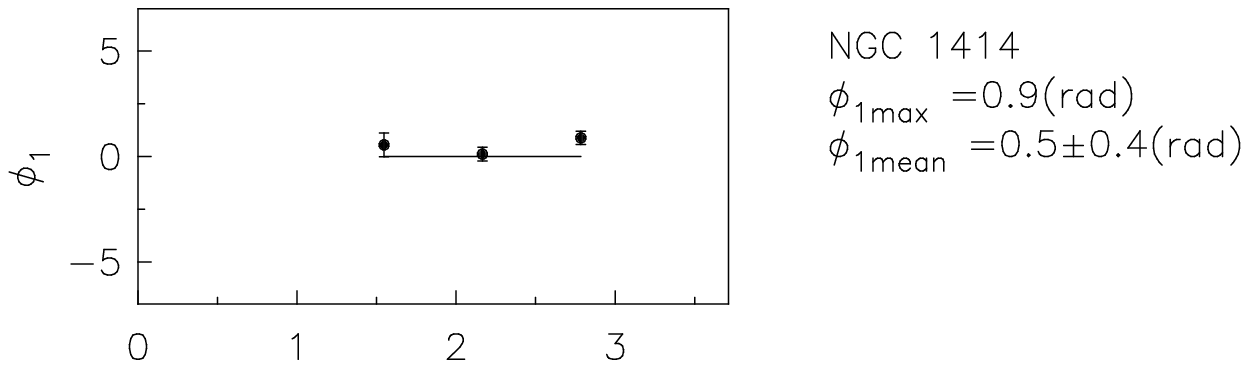}
\includegraphics[width=84mm,height=25mm]{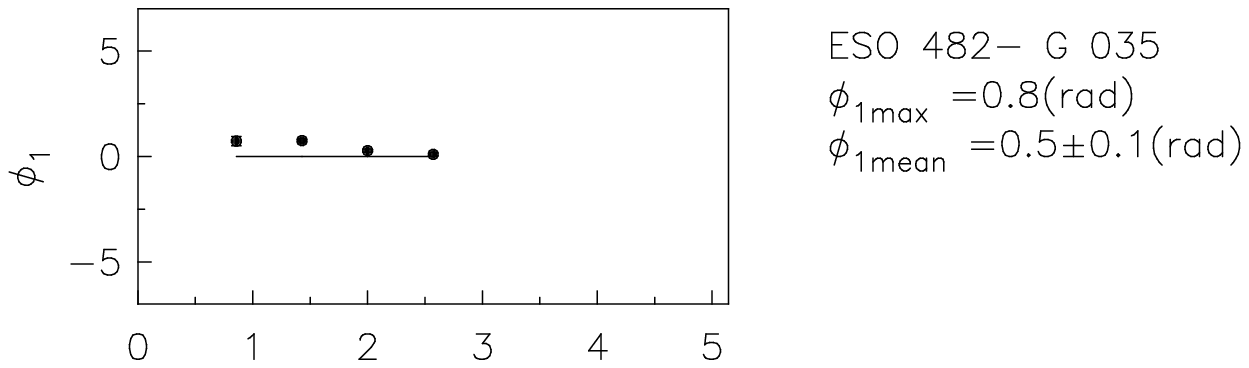}
\includegraphics[width=84mm,height=25mm]{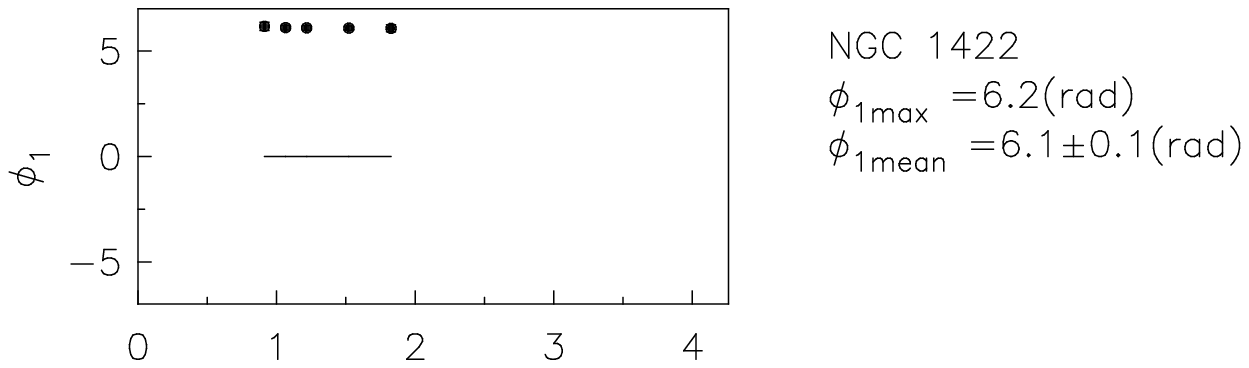}
\includegraphics[width=84mm,height=25mm]{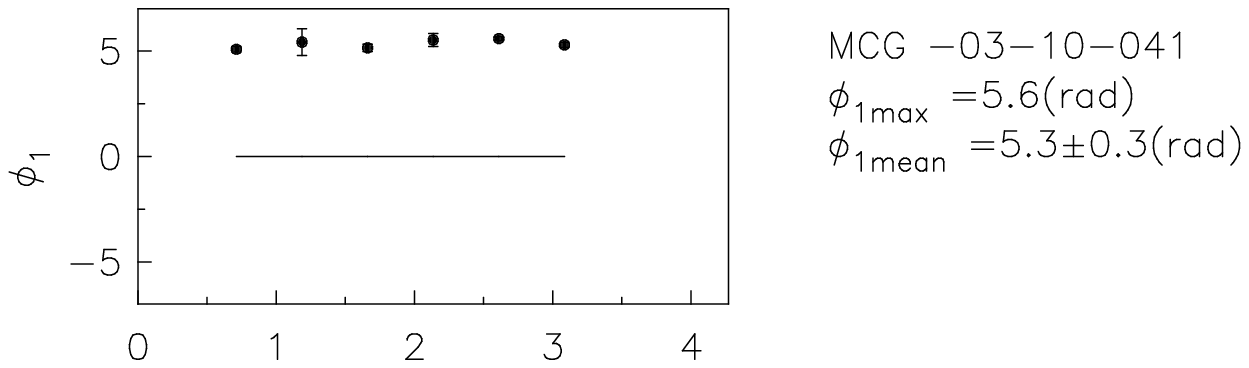}
\includegraphics[width=84mm,height=25mm]{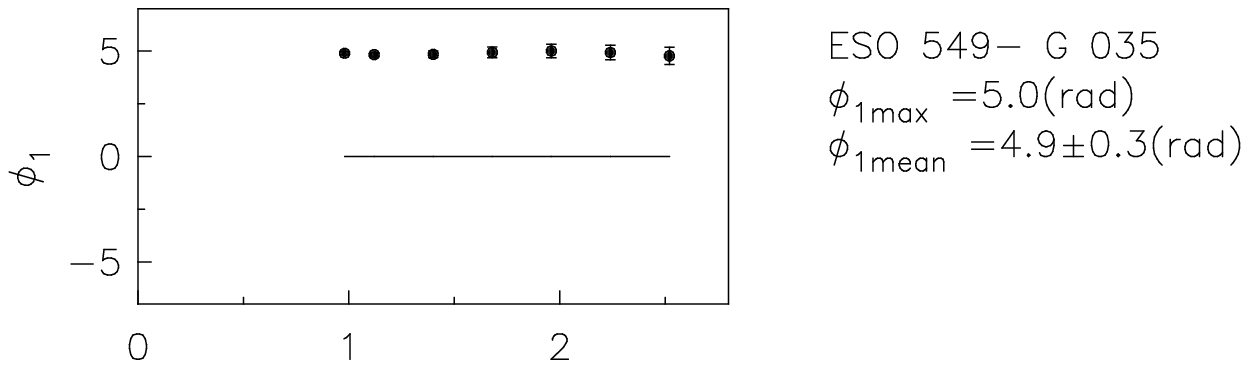}

\caption{$\phi_{1}$ coefficients of galaxies in the Eridanus group as a function of $R/R_{K}$.For NGC 1359 and ESO 548
-G 049 the J band scale length is used. For UGCA 077 \& ESO 549 -G 035, a scale length of 2kpc is used.}

\end{figure*}

\subsection{Spatial lopsidedness of the stellar component}

As a result of the HI harmonic analysis described earlier, we are at a unique position to compare the lopsidedness
observed in the stellar disc with that of the observed HI asymmetry. 
Even though a theoretical model for the 
origin of lopsidedness 
based on the linear disc response to a distorted halo predicts similar
values for the $A_1$ coefficients from the stellar and gaseous components 
(Jog 1997), this point is not yet verified by observations. Here we analyse the R-band images of some of the sample galaxies obtained from
ARIES, Nainital,India.  Details of the observations and the basic image processing are dealt with elsewhere (Omar \&
Dwarakanath 2006).

The reduced images were deprojected using the IRAF{\footnote {IRAF is distributed by the National Optical Astronomy
Observatories,which are operated by the Association of Universities for Research in Astronomy, Inc., under cooperative
agreement with the National Science Foundation.}} task IMLINTRAN (Buta et al., 1998). The mean inclination and position
angle derived from the HI velocity field were used in deprojection.Since our interest was in the outer regions of the
galaxies the bulge-disc decomposition was not performed before the deprojection.  In this region ($\sim 3$kpc from the
centre), the effect due to the bulge and the bar are unimportant.

From the deprojected images, the isophotal intensities along concentric annuli of width $1^{\prime\prime}$ were extracted as a function of azimuthal angle. The ELLIPSE{\footnote {ELLIPSE is a product of
the Space Telescope Science Institute, which is operated by AURA for NASA.} task was used for this purpose.
A $\chi^{2}$ fit on the extracted intensities was carried out by NFIT1D routine of STSDAS using the function

\begin{equation}
I(R,\phi)=a_{0}(R)+\sum_{m}a_{m}(R)\cos(m\phi)+b_{m}(R)\sin(m\phi)
\end{equation}

Here, $I(R,\phi)$ is the intensity at the ring radius $R$ and azimuthal angle $\phi$ in the plane of the galaxy. $a_{m}$ and $b_{m}$ were the
harmonic coefficients. From the resulting $a_{1}$,$b_{1}$ coefficients the normalized A$_{1}$ coefficients for various
rings were determined.

The values so derived for four galaxies: NGC 1309, NGC 1347, IC
1953, and NGC 1359, are shown in Figure 4. For easy comparison, we have 
also plotted the
corresponding $A_{1}$ coefficients derived from the HI data. It is seen that 
A$_{1}$ coefficients derived
from R-band images and those from HI analysis are comparable in the 
radial region where the data overlap, although we caution that the region 
of overlap is small. In one case, the stellar
asymmetry values are slightly higher than the HI values while the reverse 
is true
in two cases, and in NGC 1359 they overlap. Thus in general,
 the $A_1$ values for stars show the same general trend as 
do the HI values, and in the outer regions only HI is available as a tracer.
\begin{figure*}
\includegraphics[width=84mm,height=50mm]{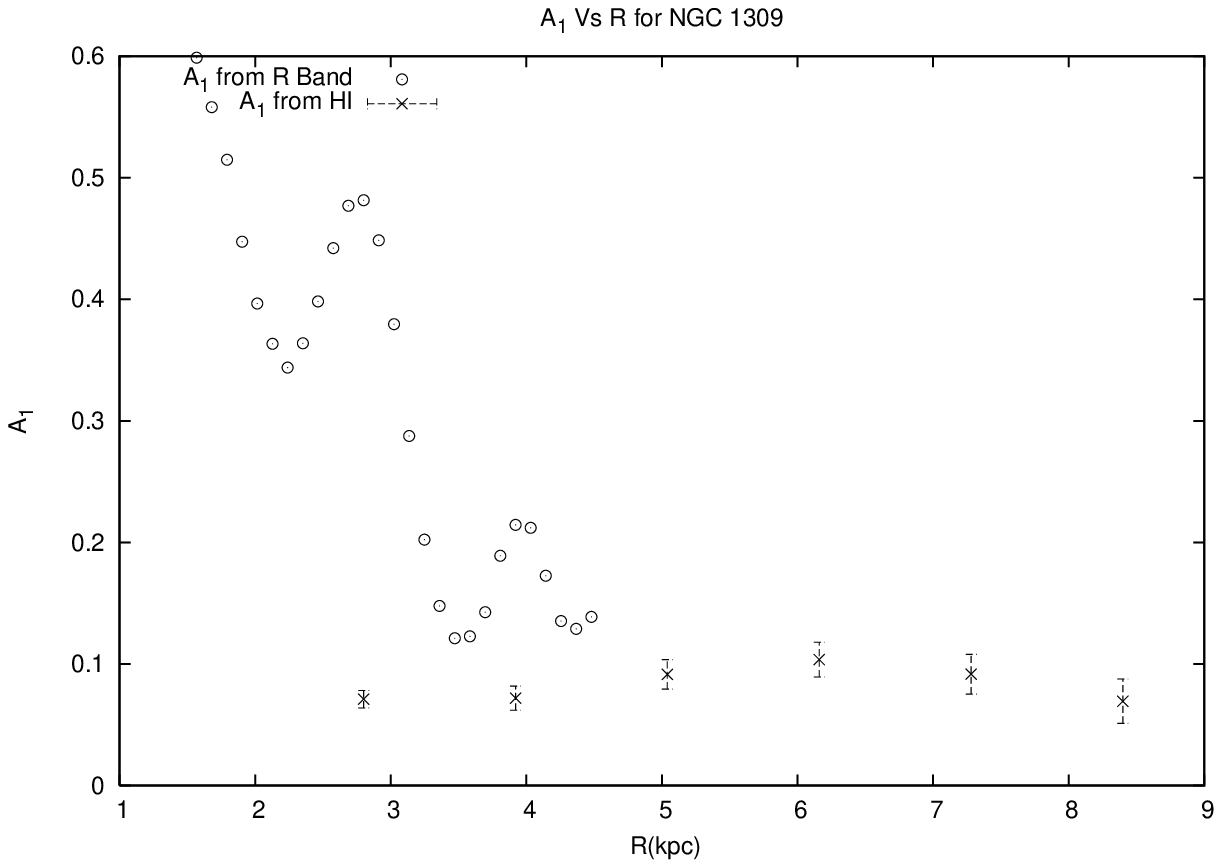}
\includegraphics[width=84mm,height=50mm]{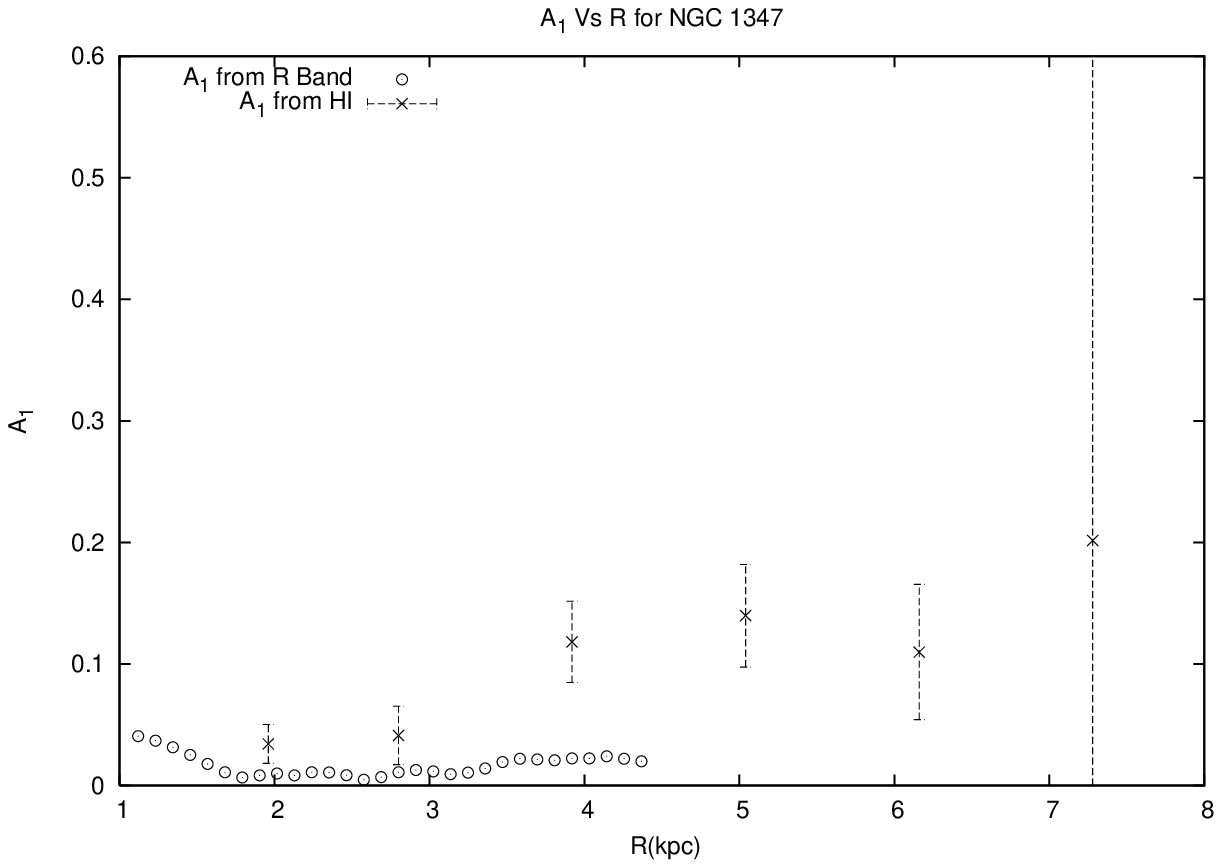}
\includegraphics[width=84mm,height=50mm]{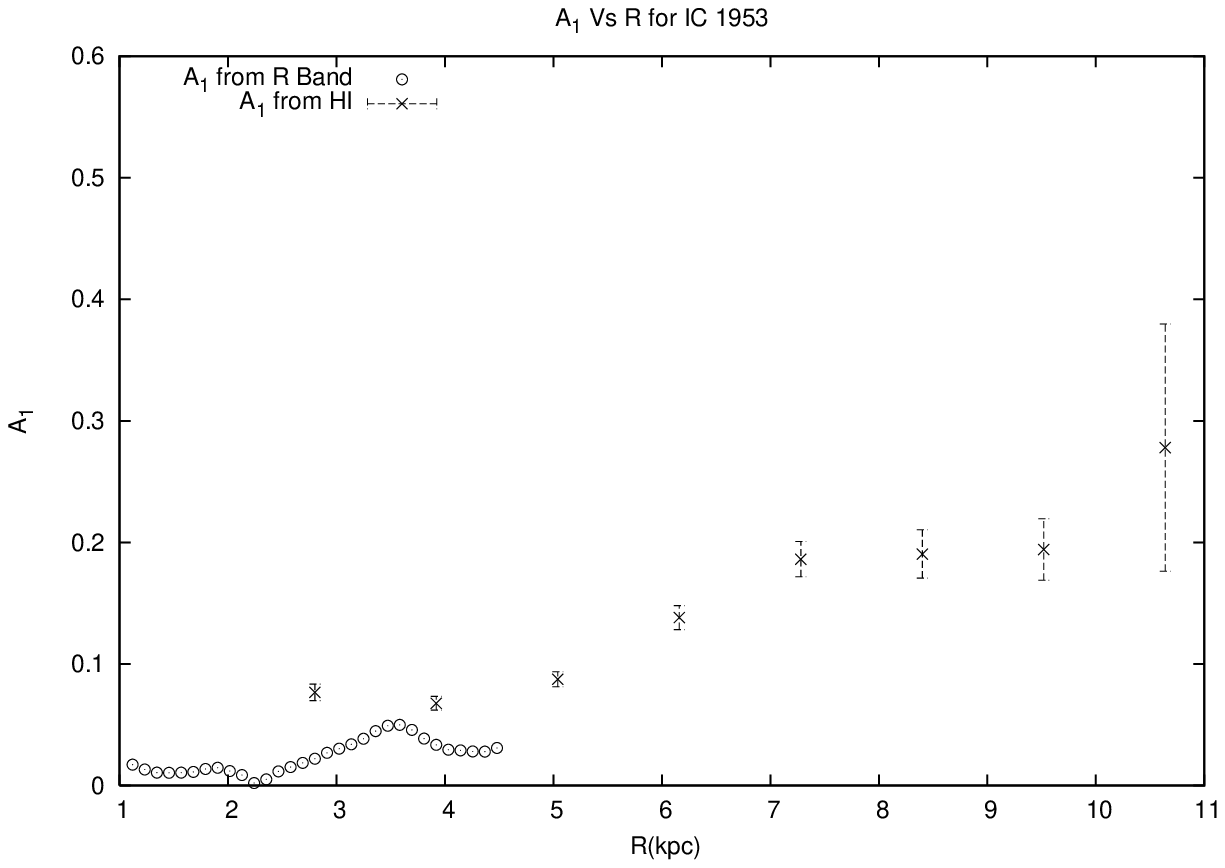}
\includegraphics[width=84mm,height=50mm]{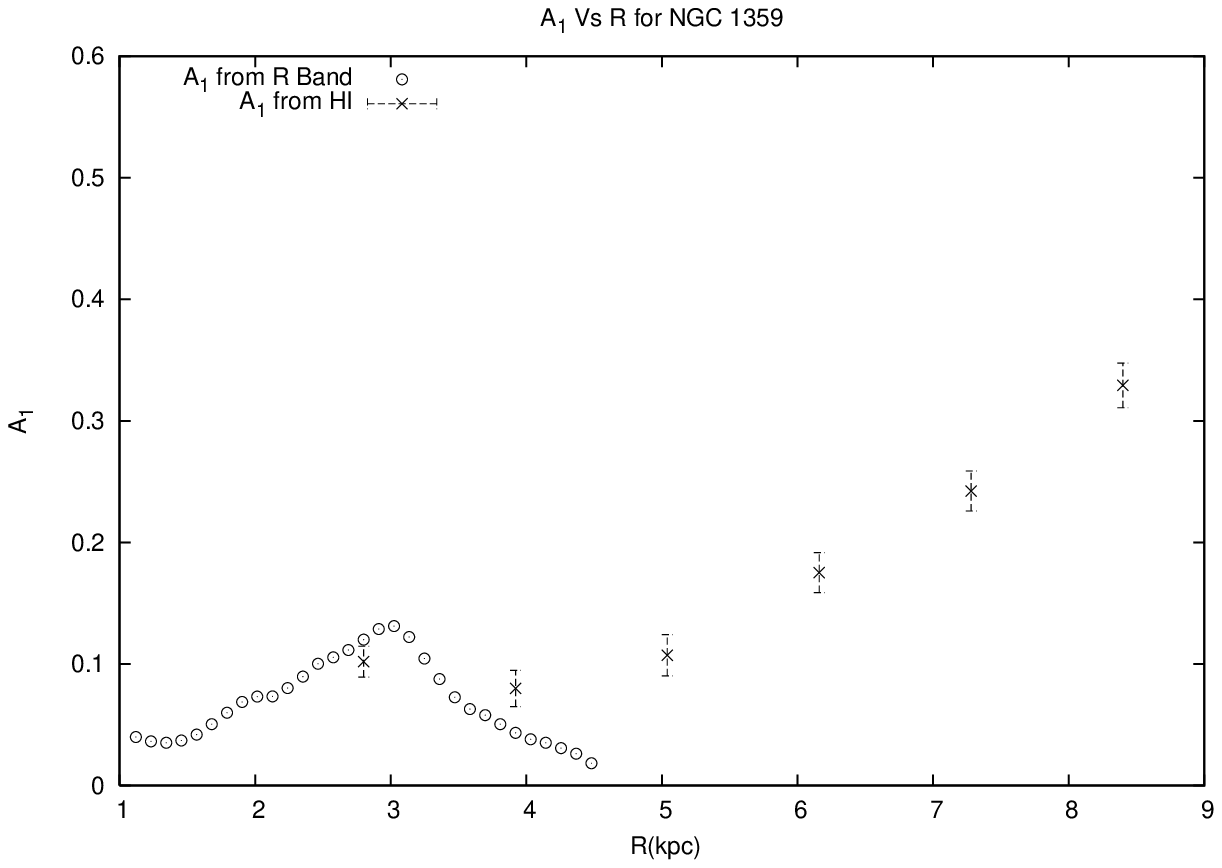}
\caption{A$_{1}$ coefficients derived from R-band images of galaxies along with
those derived from HI analysis as a function of distance.}
\end{figure*}

\subsection{Estimation of the halo perturbation}}

Assuming that the disc lopsidedness arises as a disc response to the halo perturbation, we can use the above observed
$A_{1}$ coefficients to determine the halo asymmetry or the perturbation potential(Jog, 2000). We assume that the
potential $\psi(R,\phi)$ at a radius $R$ for a galaxy to be composed of an unperturbed part $\psi_{0}(R,\phi)$ and a
perturbed part $\psi_{1}(R,\phi)$. It is known that most spiral galaxies have flat rotation curves in the outer
regions.Choosing $\psi_{0}\propto \ln(R)$ as the unperturbed potential can explain this result. The perturbation
potential is assumed to have a cosine dependence to represent the lopsidedness.

For computational purpose, $\psi_{0}$ and $\psi_{1}$ were taken as

\begin{equation}
\psi_{0}(R,\phi)=V_{c}^2 \ln (R) \nonumber
\end{equation}
\begin{equation}
\psi_{1}(R,\phi)=V_{c}^2\epsilon_{1}\cos(\phi)
\end{equation}

Here, $V_{c}$ is the rotational velocity, $\phi$ is the azimuthal angle in the plane of the galaxy and $\epsilon_{1}$
is the perturbation parameter which is assumed to be constant with radius for simplicity.

Simultaneously solving the equations of motion of orbits in this net potential, the effective surface density (assuming
an exponential disc) and the equation of continuity yields a relation between the $A_1$ values and the halo
perturbation parameter $\epsilon_1$ for an exponential disc (see Appendix, Jog, 2000). This is applicable for both stars and gas in the linear perturbation regime as shown by Jog (1997), and also observed to be true in our sample (Section 3.2).

The typical HI radial surface density profile of the galaxies belonging to Eridanus group is far from exponential and
was close to a Gaussian (Omar \& Dwarakanath 2005a). This is unlike the Virgo cluster, where many galaxies have
exponentially decreasing HI surface density in the outer regions (Warmels 1998). Hence we determined a scale length,
$R_w$, associated with a Gaussian profile for the various galaxies in this group. This was done by performing a
$\chi^2$ fit to the radial surface density profile. A face-on radial surface density profile was obtained for each
galaxy after integrating along concentric annuli. This surface density profile was fitted with a curve of the form
$S_{0}\exp(-(R-b)^{2}/2R_{w}^{2})$ using a $\chi^2$-fitting technique with $S_{0}$, $b$ and $R_{w}$ as the best fit
parameters.

Repeating the analysis as in Jog (2000), but for a Gaussian surface density distribution, we obtain the following
relation between $A_1$ and $\epsilon_1$, in terms of $R_w$ :

\begin{equation}
\epsilon_{1}=\frac{A_{1}(R)}{(2(\frac{R}{R_{w}})^{2}-1)}
\end{equation}

The values of $R_w$, the Gaussian scale length, $<A_{1}>_w$, the mean $A_1$ observed over 1-2 $R_w$ range, and
$<\epsilon_1>$, the mean perturbation parameter for the halo potential over this radial range are given in the last
three columns of Table 2. The typical value of the exponential stellar disc scale length is $\sim 2$ kpc (Table 1),
while that of the scale length for the HI distribution $R_w$ is $\sim 6$ kpc (Table 2). The sixth column gives an
average of $<A_{1}>_K$ for HI measured over 1.5-2.5 exponential disc scale lengths, and can be compared directly with the
values of lopsidedness measured earlier from stellar distribution over the same range of radii (Rix \& Zaritsky 1995,
Bournaud et al. 2005). The last two columns denote asymmetry in the HI surface density and the mean perturbation
parameter respectively, in the outer parts of a galactic disc.  The mean value of $<A_{1}>_K$ in the inner disc (1.5 to
2.5 $R_{K}$) is 0.24, while that in the outer disc is slightly higher =0.27 (Table 2).

\begin{table*}
\begin{minipage}{140mm}
\caption[]{The m=1 asymmetry values in the Eridanus sample.}
\label {Table 2}
\noindent
\begin{tabular}{@{}llccccccc@{}}
\hline
\hline

Galaxy      &   Type&Inclin(i)&    PA            &$A_{1Max}$      	&$<A_{1}>_{K}$		&$R_{w}$	&$<A_{1}>_{w}$	&$<\epsilon_{1}>$\\
	    &		&      (Deg)  & (Deg)&			&(1.5 to 2.5 $R_{K}$)	&(kpc)		&(1 to 2 $R_{w}$)	&(1 to 2 $R_{w}$)\\
\hline
	    &		&	&	&	     	  	&			&		&		&	\\
NGC 1309    &  SAbc     &        20&   210        &$0.10\pm0.01$          &$0.07$			&$4.76$		&$0.09$		&$0.04$	\\
UGCA 068    & SABcdm   &        34&    35         &$0.21\pm0.04$         	&$--$			&$3.59$		&$0.11$		&$0.04$	\\
NGC 1325    &  SAbc &        71&   232           &$0.38\pm0.01$        	&$0.18$			&$6.18$		&$0.20$		&$0.06$	\\
NGC 1345    &  SBc &        34&    88           &$0.48\pm0.04$        	&$0.09$			&$6.95$		&$0.30$		&$0.12$	\\
NGC 1347    &  SBcd&        26&   328            &$0.20\pm2.1$          	&$0.04$			&$3.23$		&$0.12$		&$0.04$	\\
UGCA 077    &  SBdm&        66&   149            &$0.10\pm0.03$        	&$0.07$			&$4.55$		&$0.06$		&$0.02$	\\
IC  1953    &   SBd&        37&   129             &$0.28\pm0.10$         	&$0.19$			&$4.61$		&$0.15$		&$0.05$	\\
NGC 1359    &  SBcm&        53&   325            &$0.33\pm0.02$         	&$0.14$			&$11.23$	&$--$ 		&$--$	\\
NGC 1371    &  SABa&        49&   136            &$0.19\pm0.01$         	&$--$			&$--$		&$--$		&$--$	\\
ESO 548 -G 049&  S?&        71&   128             &$0.80\pm0.08$         	&$0.55$			&$3.91$		&$0.74$		&$0.51$	\\
ESO 482 -G 013& Sb&        63&    65           &$0.19\pm0.03$         	&$0.15$			&$2.95$		&$0.19$		&$0.10$	\\
NGC 1385    &  SBcd&        40&   181            &$0.33\pm0.08$         	&$0.33$			&$6.16$		&$0.33$		&$0.26$	\\
NGC 1390    &  SB0/a&        60&    24            &$0.40\pm0.07$         	&$0.40$			&$4.07$		&$0.36$		&$0.18$	\\
NGC 1414    &  SBbc&        80&   357            &$0.14\pm0.04$         	&$0.12$			&$8.40$		&$--$		&$--$	\\
ESO 482-G 035&  SBab&        49&   185           &$0.83\pm0.06$	       	&$0.59$			&$3.22$		&$0.71$		&$0.26$	\\
NGC 1422    &  SBab&        80&    65            &$0.79\pm0.10$         	&$0.58$			&$20.45$	&$--$		&$--$	\\
MCG -03-10-041&  SBdm&        57&   343          &$0.31\pm0.05$         	&$0.17$			&$5.15$		&$0.28$		&$0.12$	\\
ESO 549-G 035&  Sc&        56&    30            &$0.17\pm0.02$         	&$0.09$			&$3.73$		&$0.09$		&$0.05$	\\
\hline
\hline
\end{tabular}

{\it Notes:}The
$A_{1}$ mean value between 1.5 to 2.5 exponential scale lengths is given in column 6. The Gaussian scale length for HI, $R_{w}$, and the resulting $<A_{1}>_w$ and $\epsilon_{1}$ over 1 - 2 $R_{w}$ are shown in columns 7 and 8 respectively.\\The mean of $<A_{1}>_K$ in the range 1.5 to 2.5 $R_{K}$ is $0.24\pm0.19$ and the mean of $<A_{1}>_w$ in the 1 to 2 $R_{w}$ range is $0.27\pm0.22$.The mean of $<\epsilon_{1}>$ is $0.13\pm 0.13$.\\ These results do not change much if the galaxies with i $> 70^\circ$ and with $R_{exp}=2$ are eliminated from the sample - in that case the mean of $<A_{1}>_{K} =0.22\pm0.17$ between 1.5 - 2.5 $R_K$ and mean of $<A_{1}>_{w}=0.26\pm0.18$ between 1-2 $R_w$ 

\end{minipage}
\end{table*}

\section{Discussion}

\noindent {\bf {1. Distribution of lopsidedness}}

We have carried out the Fourier harmonic analysis for the HI surface density of the Eridanus group of galaxies. All the
eighteen galaxies studied show significant average lopsidedness with a mean value of $<A_{1}>_K$ = 0.24 in the inner
regions of $< 5$ kpc, which is more than twice the average value observed for field galaxies (Zaritsky \& Rix 1997,
Bournaud et al. 2005). A large fraction $\sim 30\%$ show even higher lopsidedness with a value of $<A_{1}>_K\geq0.3$,
whereas only 7\% of the field galaxies have such high lopsidedness (Bournaud et al. 2005). In the field galaxies $\sim 12\%$
of galaxies show $A_1 \geq 0.2$, whereas in the Eridanus sample $\sim 40\% $ of the galaxies show this.In the present
paper, we have measured the values in the outer discs ($> $ 5 kpc) or outside of 2.5 exponential disc scale lengths as
well, and find that the average value of mean lopsidedness measured in the outer regions is slightly higher(= 0.27).

\medskip

\noindent {\bf {2. Phases of lopsidedness}}

From Fig. 3, we see that the values of the phase angles of 15 galaxies in the Eridanus group remains nearly constant
without sudden jumps; the exceptions being NGC 1309, UGCA 077 and NGC 1371.This means the surface density contours have
egg-shaped rather than one-armed profiles, and the potential causing the disc lopsidedness can be taken to have no
radial phase dependence. This is similar to the behaviour seen in the inner regions as traced by the near-IR studies in
Rix \& Zaritsky (1995)- see Jog (1997) for a discussion of this topic.
A nearly constant phase implies that these are global m=1 modes.

\noindent {\bf {3.  Origin of disc lopsidedness in group galaxies}}

We have shown above that the overall average values of $A_1$ in the Eridanus group galaxies in the inner regions are
higher by a factor of two compared to the field galaxies(\mbox{$<A_{1}>$}$=0.11$,Bournaud et al.,2005), and $\sim 70\%$ of the
sample galaxies show such a high value of lopsidedness. The
similar values for lopsidedness measured in both stars and gas
in the inner regions show that this indicate true
lopsidedness. However we caution that, since the number of galaxies studied in HI and R-band is small and the radial range of overlap for the comparison (see Fig. 4) is small, some of the
difference in the lopsidedness in the group vs. the field cases could 
perhaps still be 
attributed to the different tracers used (HI for the group case
and the stars for the field case respectively.) 
 The overall higher value of $A_1$ measured implies that a 
group environment is more effective in
generating lopsidedness in discs of galaxies, either via tidal interactions 
that can distort the halo and then affect
the disc (Jog 1997), or via asymmetric gas accretion (Bournaud
et al. 2005). While we cannot give a clear preference for either one of these mechanisms
based on the present work, tidal interactions appear to be more likely as argued below.

Given the high number density of galaxies in a group, a higher frequency and strength of tidal interactions are
expected. Thus a tidal origin can explain the high frequency as well as the higher strength of lopsidedness observed in
the Eridanus group of galaxies compared to a sample of field galaxies.

Theoretically one can explain the similar observed values of
lopsidedness  for stars and HI gas 
(Section 3.2), if the origin of lopsidedness is due to a linear disc 
response to a
distorted halo (Jog 1997). In this case, the likely origin of
the halo distortion or lopsidedness could be due to tidal interactions between
galaxies (Weinberg 1995).

It is interesting to note that the galaxies in the Eridanus 
group exhibit HI deficiency, which is ascribed to tidal
interaction (Omar \& Dwarakanath 2005b). This also might indicate that the higher average values of lopsidedness which
we have observed could be due to tidal interaction.

We note that in contrast, a typical field sample showed no
correlation in the stellar lopsidedness measured in the inner
disc regions with
a tidal parameter (Bournaud et al. 2005).

In the field case, the late type galaxies show a higher
lopsidedness and are also more likely to be lopsided (see Fig.
7, Bournaud et al. 2005; also see Zaritsky \& Rix 1997, Matthews et al. 1998). 
To check if this could be a spurious
reason for the high $A_1$ measured in our sample, We plotted $A_1$ vs.
the galaxy type for our sample. Interestingly, this shows an
opposite
trend, namely we get a weak correlation showing a {\it decrease}
in $A_1$ for later type galaxies (Figure 5). This is in sharp contrast
to the strong correlation in $A_1$ with galaxy type, with $A_1$
increasing for later type galaxies, that is seen in previous
studies which involved field galaxies.

\begin{figure*}
\includegraphics[width=84mm,height=60mm]{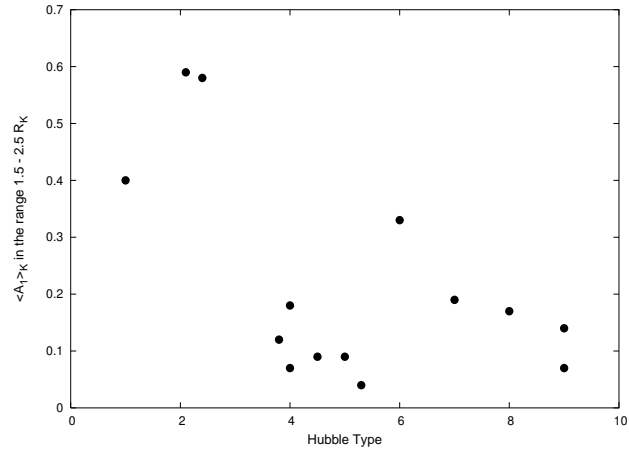}
\caption{$<A_{1}>_K$ in the 1.5 -2.5 $R_K$ range Vs. Hubble type.The lopsidedness decreases for late-type galaxies. This is opposite to the trend seen in the field galaxies.}
\end{figure*}
Hence the high values of $A_1$ measured here
cannot be due to the type of galaxies included in the sample.

These two results: namely, the higher average value of $A_1$ measured 
for the Eridanus group galaxies, and the weak anti-correlation of $A_1$ vs. 
galaxy Hubble type, clearly indicate that the main physical mechanism for the 
origin of the disc lopsidedness in a group environment is different from that 
for the field galaxies. Perhaps the gas accretion, which plays an important 
role in causing lopsidedness  in the field galaxies (Bournaud et al. 2005), 
may not be so important in a group, especially since a group
probably does not have much cold gas.

We note that the earlier work on the measurement of HI asymmetry in Sculptor group galaxies (Schoenmakers 2000) also
showed kinematical lopsidedness in all the five galaxies studied. However, this method gives a value for the
perturbation parameter for the potential times a term dependent on the inclination angle only. The average value for
this product is $\sim 0.06$ which is much smaller than the average value of $<\epsilon_1>$= $0.13\pm0.13$ (Table 2)
obtained in the present paper for the 18 Eridanus group galaxies.

\section{Conclusion}

 We have measured $A_1$ the amplitude of the first Fourier component 
 over the average value for the surface density of
HI in a sample of 18 galaxies in the Eridanus group of galaxies, 
by Fourier analysis of the interferometric 2-D data
(Omar \& Dwarakanath 2005a) on these galaxies.
This is the first quantitative measurement of the
spatial lopsidedness using HI data. 
The summary of conclusions from this paper is as follows:

1. All the galaxies studied show significant lopsidedness, with average 
$A_1 > 0.2$ in the region of 1.5-2.5 disc scale lengths. 
 A large fraction $\sim 30 \%$ show
even higher average lopsidedness $(> 0.3)$.
 For a few of the galaxies, the stellar R-band data available in the inner
regions were analyzed, and the resulting values of lopsidedness are
shown to be similar to the HI lopsidedness.
The same amplitudes for stars and gas 
can be naturally explained
if both arise due to a linear response of the disc to a
distorted or lopsided halo (Jog 1997).

2. The lopsidedness is observed to increase with radius, and the outer regions have an average $A_1$ value of $\sim
0.27$.

3. The present work measures $A_1$ in discs up to the edge of the optical discs or four exponential disc scale lengths,
and in a few cases even beyond that. This is more than twice the distance that is typically studied in the stellar
distribution via near-IR photometry.  This can help provide constraints on the origin of lopsidedness in discs,
especially since lopsidedness is higher at larger radii.From the observed $A_1$ values, the halo distortion is deduced
to be $\sim 10 \% $.

4.  The overall higher value of lopsidedness
$A_1$ measured in the inner regions in the Eridanus group galaxies 
compared to the field galaxies (e.g., Bournaud et al.
2005); and the smaller values of lopsidedness observed for the later 
Hubble type galaxies - which is opposite to the trend seen 
in field galaxies,
together imply that a different physical mechanism is
responsible for the origin of the disc lopsidedness in a
group environment.

\bigskip

The present work highlights the need for a future dynamical study of the origin and evolution of disc lopsidedness in
galaxies in groups. This can help in understanding the interactions and also the halo properties for galaxies in
groups.

\section{ACKNOWLEDGMENTS}

We thank the referee, Frederic Bournaud, for critical comments especially
stressing the necessity of making a  comparison of the stellar and HI
lopsidedness to confirm that the origin of the measured asymmetry in HI
is due to the group environment. We thank Ron Buta and Aparna
Chitre for useful email 
correspondence regarding the near-IR data analysis. This work is partially
supported by the grant IFCPAR/2704-1.

R.A.Angiras takes great pleasure in thanking K. Indulekha of School of Pure and Applied
Physics, M.G.University for her help and encouragement during this project.He also thanks the University Grants
Commission of India and St.Joseph's College, Bangalore for granting study leave under the FIP leave of $10^{th}$ Five
Year Plan and Raman Research Institute, Bangalore for providing all the facilities to pursue this study.

\end{document}